\documentclass[12pt]{article}
\usepackage{amssymb}
\usepackage{amsmath}
\usepackage{amstext}
\usepackage{float}

\usepackage{verbatim} 
\usepackage{fancyhdr}
\usepackage{fancybox}
\usepackage{color}
\usepackage{ulem,bbold}
\usepackage{enumitem}
\usepackage{subfigure}
\usepackage{bbm}
\usepackage{parskip}
\usepackage{cite}
\usepackage{xcolor}
\usepackage{graphicx}

\usepackage{makecell}
\usepackage{array}

\newcommand{\Comment}[1]{{}}
\definecolor{MyDarkBlue}{rgb}{0.15,0.15,0.45}
\usepackage[linktocpage=true]{hyperref}
\hypersetup{
colorlinks=true,
citecolor=MyDarkBlue,
linkcolor=MyDarkBlue,
urlcolor=MyDarkBlue,
}

\newcommand{\be}{\begin{equation}}
\newcommand{\ee}{\end{equation}}
\newcommand{\bea}{\begin{eqnarray}}
\newcommand{\eea}{\end{eqnarray}}
\newcommand{\beas}{\begin{eqnarray*}}
\newcommand{\eeas}{\end{eqnarray*}}
\newcommand{\nn}{\nonumber}

\newcommand{\llp}{\left [}
\newcommand{\rrp}{\right ]}
\newcommand{\lp}{\left (}
\newcommand{\rp}{\right )}

\newcommand{\dd}{\text{d}}

\numberwithin{equation}{section}

\title{Radial solutions de sitter}

\begin{document}

\begin{center}
{\Large \bf{Radial Solutions of Multi-Field de Sitter Galileons\\}}
\vspace{1cm}
\end{center}

\vspace{1truecm}
\thispagestyle{empty}
\centerline{\Large Alice Garoffolo,${}^{\rm a,}$\footnote{\href{mailto:aligaro@sas.upenn.edu}{\texttt{aligaro@sas.upenn.edu}}} Mary Gerhardinger,${}^{\rm a,}$\footnote{\href{mailto:maryge@sas.upenn.edu}{\texttt{maryge@sas.upenn.edu}}} Kurt Hinterbichler,${}^{\rm b,}$\footnote{\href{mailto:kurt.hinterbichler@case.edu} {\texttt{kurt.hinterbichler@case.edu}}} }

\vspace{0.5truecm}
\centerline{\Large Mark Trodden,${}^{\rm a,}$\footnote{\href{mailto:trodden@upenn.edu} {\texttt{trodden@upenn.edu}}} }

\vspace{.5cm}

\centerline{{\it ${}^{\rm a}$Center for Particle Cosmology, Department of Physics and Astronomy,}}
\centerline{{\it University of Pennsylvania, Philadelphia, Pennsylvania 19104, USA }}
\vspace{.25cm}

\centerline{{\it ${}^{\rm b}$CERCA, Department of Physics,}}
\centerline{{\it Case Western Reserve University, 10900 Euclid Ave, Cleveland, OH 44106}} 
\vspace{.25cm}

\begin{abstract}
We study static, spherically symmetric screening solutions in the $\mathfrak{so}(N)$-invariant multi-field de Sitter Galileon theory, with matter couplings preserving the internal symmetry. Unlike in flat space, the de Sitter radial problem is affected both by the cosmological horizon and by the possible appearance of a finite strong-coupling radius, where the radial equation becomes singular and the effective description breaks down. 
Near the source, the quartic interaction governs the nonlinear screened branch, while farther away the quadratic terms dominate, defining a de Sitter analog of the Vainshtein radius at their crossover. A screened solution is viable only if this crossover occurs before the strong-coupling radius is reached. We  study perturbations around the radial background and identify branches for which perturbations are stable and subluminal, within a restricted radial domain set by the strong-coupling point.
Finally, we compare the de Sitter and Anti-de Sitter cases, isolating effects due to curvature from those specific to the cosmological horizon. 
Our results indicate that curvature can alleviate the superluminality issues that arise in Galileon theories for appropriate choices of matter couplings, though this comes at the price of a finite region of validity for the effective theory.
\end{abstract}

\newpage
\tableofcontents

\section{Introduction}
One of the central problems that infrared modifications of gravity face is how to reconcile additional light degrees of freedom with the empirical successes of general relativity near localized sources. In most constructions, the new propagating mode is a light scalar, which, once coupled to matter, typically mediates a long range fifth force. Any viable theory must therefore contain a {\it screening mechanism} that suppresses this force close to matter while allowing departures from general relativity on large scales. Among the known possibilities, the Vainshtein mechanism \cite{Vainshtein:1972sx} is distinctive in that the suppression is driven by derivative self-interactions: the same nonlinearities responsible for modifying gravity at large distances also restore agreement with general relativity near sources (see \cite{Babichev:2013usa} for a review). 
A basic issue is not only whether screening solutions exist, but whether they remain perturbatively viable once one considers stability, strong coupling, and propagation speeds around them.

Galileon theories provide a particularly sharp arena in which to study these questions. They first aroused wide interest due to their appearance in the decoupling limit of Dvali-Gabadadze-Porrati gravity and related infrared modifications of gravity~\cite{Dvali:2000hr,Dvali:2000xg,Luty:2003vm,Nicolis:2004qq}, and then more generally as effective field theories with derivative self-interactions arranged so as to preserve second-order equations of motion~\cite{Nicolis:2008in,Deffayet:2009wt,Deffayet:2009mn,Nicolis:2009qm}. Their nonlinear structure naturally realizes Vainshtein screening and has made them a standard framework in studies of modified gravity and cosmology~\cite{Chow:2009fm,Silva:2009km,Babichev:2009jt,Kobayashi:2009wr,Khoury:2009tk,Kobayashi:2010wa,Wyman:2010jp,Creminelli:2010ba,DeFelice:2010pv,Babichev:2010jd,DeFelice:2010nf,Kimura:2011dc,Sjors:2011iv,Chkareuli:2011te,Sbisa:2012zk,deRham:2014zqa,Gumrukcuoglu:2021gua,Goon:2011xf,Burrage:2010cu,Andrews:2013qva,Andrews:2013uca,Kase:2015zva,Hinterbichler:2016try,Gerhardinger:2022bcw,Burrage:2023eol,SevillanoMunoz:2024ayh,Gerhardinger:2024rza,Garcia-Saenz:2025htt}. At the same time, these theories have long-faced several challenges: radial screening solutions are often plagued by instabilities, and perturbations around them frequently exhibit superluminal propagation speeds.

In the single-field case, superluminality around radial backgrounds has long been recognized as a generic feature. In DGP and related Galileon theories, superluminal propagation on Lorentz-breaking backgrounds was established early on~\cite{Nicolis:2008in,Hinterbichler:2009kq,Goon:2010xh}. While the appearance of superluminality does not necessarily imply an immediate breakdown of causality at the level of the effective theory \cite{Burrage:2011cr,Goon:2016une}, it may obstruct a standard Lorentz-invariant UV completion, as it is connected to analyticity and positivity arguments~\cite{Adams:2006sv,deRham:2017avq,Burrage:2020bxp}.
The multi-field case is still more constrained. Multi-Galileon theories arise naturally in higher-codimension brane constructions, where several brane bending modes appear simultaneously and the four-dimensional theory inherits an internal symmetry acting on the scalar multiplet~\cite{Dubovsky:2002jm,Gabadadze:2003ck,Kolanovic:2003am,deRham:2007xp,deRham:2007rw,Kaloper:2007ap,Corradini:2007cz,Corradini:2008tu,deRham:2010rw,Padilla:2010de,Deffayet:2010zh,Padilla:2010ir,Hinterbichler:2010xn,Burrage:2011cr,Goon:2014ywa}. Static spherically symmetric solutions in these theories have been studied in flat space both in the bi-Galileon case and in the more general $\mathfrak{so}(N)$-symmetric setting~\cite{Andrews:2010km,Padilla:2010tj,Padilla:2010de,deFromont:2013iwa}. Those analyses showed that screened radial branches persist, but the instability and superluminality do so as well, arguing that the only possible way to have stable screening solutions is by introducing more complicated couplings or allowing non-trivial asymptotic behavior of the fields. 
The tension between screening, stability, and acceptable propagation speeds is therefore one of the structural issues of this subject.

In this paper, we show that the issues of stability and superluminality around spherically symmetric screening solutions can be alleviated when curvature effects are taken into account. In particular, we consider the multi-field generalization \cite{Garoffolo:2025igz} of de Sitter Galileons~\cite{Burrage:2011bt,Goon:2011qf,Goon:2011uw}, which extends the flat-space $N$-field multi-Galileon theory with internal $\mathfrak{so}(N)$ symmetry to a de Sitter background. 
We find static, spherically symmetric configurations sourced by an $\mathfrak{so}(N)$ symmetry preserving a non-derivative coupling to matter and ask under what conditions the theory admits screened radial solutions, how horizon regularity restricts the allowed branches, and whether there exist solutions for which perturbations remain stable, non-superluminal and weakly coupled in the screened region.
We find that a distinctive feature of the de Sitter problem is that, in addition to the usual Vainshtein radius $r_{\rm V}$, where non-linear effects and screening become important, curvature introduces an extra length scale $r_{\rm SC}$, that we call the strong coupling radius, at which the radial equation becomes singular. This occurs when the coefficient of the second radial derivative in the radial equation vanishes, signaling the onset of strong coupling and the breakdown of the effective description. The viability of screening in de Sitter space is therefore controlled by the relative order between the Vainshtein radius and this strong-coupling radius.

We then analyze perturbations around the screened backgrounds. We find that, for certain choices of boundary conditions, there exist solutions in which perturbations are both stable and subluminal, but only up to the strong-coupling radius, rather than extending all the way to the horizon.
Finally, we compare the de Sitter case, which has a cosmological horizon, to the Anti-de Sitter case, which does not.  This helps disentangle effects due to curvature from those specifically associated with the presence of a cosmological horizon.

The paper is organized as follows. In Section~\ref{sec:Thoery}, we present the de Sitter multi-Galileon theory with a symmetry-preserving matter coupling and discuss its symmetries. In Section~\ref{sec:radial}, we construct static, spherically symmetric solutions and identify both the de Sitter analog of the Vainshtein radius and the strong-coupling radius that can obstruct a global screened profile. In Section~\ref{sec:perturbation}, we study perturbations around these backgrounds and analyze their stability. In Section~\ref{sec:AdS} we compare the de Sitter problem with its Anti-de Sitter counterpart. Technical results are collected in the appendices.

\section{Theoretical set-up}\label{sec:Thoery}

We consider the de Sitter (dS) extension of the multi-field Galileon theory, introduced in~\cite{Garoffolo:2025igz}. The theory contains $N$ real scalar fields $\pi^I$, with $I=1,\ldots,N$, propagating on a fixed four-dimensional dS background with Hubble parameter $H$.  This theory arises from the small field limit of the theory of a dS$_4$ brane probing a dS$_{4+N}$ bulk, with an $SO(N)$ symmetry around the brane.  It is a multi-field generalization of the single-field dS Galileon derived in~\cite{Burrage:2011bt,Goon:2011qf,Goon:2011uw}, building off of \cite{deRham:2010eu} (there are simple cases of a much wider class of shift symmetric theories on dS and AdS \cite{Bonifacio:2018zex,Bonifacio:2019hrj,Bonifacio:2021mrf,Hinterbichler:2022vcc,Bonifacio:2023prb}, which govern the longitudinal mode in the partially massless limit of massive gravity on dS \cite{DeRham:2018axr}), and a dS generalization of the flat space multi-Galileon theory of \cite{Hinterbichler:2010xn}.

In four dimensions, the dS Galileon Lagrangian truncates at quartic order, and the dynamics are governed by
\be\label{eq:Action}
{\cal S}=\int \dd^4x\,\lp {\cal L}_{(2)}+\lambda\,{\cal L}_{(4)}\rp \,,
\ee
where $\lambda$ has mass dimension $[\mathrm{mass}]^{-6}$. The quadratic and quartic terms are 
\begin{align}
    \frac{1}{\sqrt{-g}}\,{\cal L}_{(2)} &= - \frac12 \,\pi^{I\mu}\pi_{I\mu}+ 2 H^2\pi^2\,, 
\label{eq:LagrangianL2}\\
    \frac{1}{\sqrt{-g}}\,{\cal L}_{(4)} &= \frac12\pi_I^{\ \mu}\pi_J^{\ \nu} \lp \pi^{I\rho}_{\ \ \nu}\pi^J_{\ \rho\mu}-\pi^I_{\ \mu\nu}\Box\pi^J\rp \nonumber\\
    &\quad + \frac{H^2}{2}\lp -2\pi^{I\mu}\pi_I^\nu\,\pi_J\pi^J_{\ \mu\nu} +2\pi^{I\mu}\pi_{I\mu}\,\pi_J\Box\pi^J
    -\frac12\pi^{I\mu}\pi_I^\nu\pi^J_\mu\pi_{J\nu}
    +\frac54\,\pi^{I\mu}\pi_{I\mu}\pi^{J\nu}\pi_{J\nu}\rp \nonumber\\
    &\quad + \frac{3 H^4}{2}\lp \pi^2\pi^{I\mu}\pi_{I\mu}
    +2\pi^I\pi^J\pi_I^{\mu}\pi_{J\mu}\rp 
    -3 H^6\pi^4\,,
\label{eq:LagrangianL4}
\end{align}
where $\pi^{I\mu}\equiv \nabla^\mu\pi^I$, $\pi^{I \mu \nu} \equiv \nabla^\mu \nabla^\nu\pi^I$ and $\pi^2\equiv\pi^I\pi_I$. Spacetime indices are raised and lowered with the dS metric $g_{\mu\nu}$, while internal indices are raised and lowered with $\delta_{IJ}$. $\nabla$ is the dS Levi-Civita connection, $\Box\equiv\nabla_\mu\nabla^\mu$, and $g$ is the determinant of the dS metric. 
The covariant equations of motion of the action~\eqref{eq:Action} are given by
\begin{align} \label{eq:EOMCovariant}
    0&=\Box\pi^K \llp  \frac{2}{\lambda}+\pi^{I}_{\mu\nu}\pi_{I}^{\mu\nu} -\Box\pi^I\Box\pi_I-4H^2\pi^I\Box\pi_I -6H^4\pi^I\pi_I \rrp \nonumber\\
    &+\pi^{K\mu\nu} \llp  2\lp \pi^{I}_{\mu\nu}\Box\pi_I -\pi^{I}_{\mu\alpha}\pi_{I\nu}^{\ \ \alpha}\rp  +4H^2\pi^I\pi_{I\mu\nu} \rrp \nonumber\\
    &+\pi^K \llp  \frac{8H^2}{\lambda} +2H^2\lp \pi^{I}_{\mu\nu}\pi_{I}^{\mu\nu} -\Box\pi^I\Box\pi_I\rp  -12H^4\pi^I\Box\pi_{I} -24H^6\pi^I\pi_I\rrp \,.
\end{align}

We will be interested in static, spherically symmetric configurations, so we work in static patch coordinates, where the metric reads
\be\label{eq:MetricStaticPatch}
    \dd s^2=-\lp 1-H^2r^2\rp \dd t^2+\frac{\dd r^2}{1-H^2r^2}+r^2\dd\Omega_2^2\,,
\ee
with $\dd \Omega_2^2=\dd \theta^2+\sin^2\theta  \, \dd\phi^2$.
Here $r\in (0,1/H)$, with $r=0$ the origin and $r=1/H$ the static patch horizon.

\subsection{Symmetries}
\label{sec:symmetries}

Other than the free coefficient $\lambda$, the action~\eqref{eq:Action} is entirely fixed by symmetry.  There are three kinds of symmetries: First, since the action is covariant on four-dimensional dS, it is invariant under the spacetime isometry algebra $\mathfrak{so}(1,4)$, under which the fields transform linearly. Second, the fields transform linearly in the vector representation of a global internal $\mathfrak{so}(N)$ symmetry,
\be\label{eq:soSymmetry}
\delta_{\Omega_{JK}}\pi^I=\pi_J\,\delta^I_{\ K}-\pi_K\,\delta^I_{\ J}\,.
\ee
Third, the action is invariant under non-linearly realized dS Galileon shifts, which are the curved-space analog of the flat-space Galilean symmetry. There are $5N$ of these symmetries: using dS in static patch coordinates~\eqref{eq:MetricStaticPatch}, they take the form
\begin{align} 
\delta_{\Omega^+_J}\pi^I &=
\sqrt{1-H^2r^2}\,\sinh(Ht)\,\delta^I_{\ J}\,, \nn\\
\delta_{\Omega^-_J}\pi^I &=
\sqrt{1-H^2r^2}\,\cosh(Ht)\,\delta^I_{\ J}\,, \nn\\
\delta_{\Omega_{rJ}}\pi^I &=
rH\sin\theta\cos\phi\,\delta^I_{\ J}\,,
\label{eq:GalileonShiftsStatic}\\
\delta_{\Omega_{\theta J}}\pi^I &=
rH\sin\theta\sin\phi\,\delta^I_{\ J}\,, \nn\\
\delta_{\Omega_{\phi J}}\pi^I &=
rH(\sin\theta\cos\phi-\cos\theta)\,\delta^I_{\ J}\,. \nn
\end{align}
These transformation laws can be found from the small field limit of the ones presented in~\cite{Garoffolo:2025igz}, after changing coordinates. 
Together, the dS isometries, internal rotations, and nonlinear Galileon shifts close to form an algebra which is a contraction of the $\mathfrak{so} (1,4+N)$ algebra of the brane worldvolume theory of~\cite{Garoffolo:2025igz}, where the contraction comes from taking the small field limit, analogously to the flat space case \cite{Goon:2012dy}.

\section{Screening solutions}
\label{sec:radial}

We are interested in studying field profiles around a localized, static, spherically symmetric source, and we therefore couple the multi-Galileon fields to a point source with mass $M$. Following~\cite{Andrews:2010km}, we consider the most general non-derivative coupling compatible with the internal $\mathfrak{so}(N)$ symmetry,
\be\label{eq:coupling}
    \frac{1}{\sqrt{-g}}{\cal L}_{\rm coupling}=- \frac{M }{2}\,P(\pi^2) \frac{\delta^{(3)}(\mathbf{x})}{\sqrt{\gamma}}\,,
\ee
where $P(\pi^2)$ is a dimensionless arbitrary function of the invariant $\pi^2\equiv \pi^I\pi_I$, and $\gamma$ is the determinant of the 3D metric, such that $\sqrt{- g } = \sqrt{g_{tt}} \sqrt{\gamma} $.
Similarly to the analysis carried out in the flat case~\cite{Andrews:2010km}, we look for radial solutions of the form
\be
    \pi^I=\delta^I_{\ N}\,\pi(r)\,, \label{piansatze}
\ee
so that the multi-field Eq.~\eqref{eq:EOMCovariant} on the static patch Eq.~\eqref{eq:MetricStaticPatch} reduces to a single equation for the radial profile $\pi(r)$. 
This equation is
\begin{align}\label{eq:RadialEOM}
    B(\pi,\pi',\lambda, r)\,\pi''+A(\pi,\pi',\lambda,r) =-\frac{M_\pi}{2\pi}\,\delta(r)\sqrt{1-H^2r^2}\,,
\end{align}
where a prime indicates a radial partial derivative, and where we have defined
\be\label{eq:Mpi}
M_\pi \equiv M \: \pi(0) \: \frac{\partial P}{\partial \pi^2} \Bigg|_{\pi(0)}\,,
\ee
which is the dimensionless effective scalar charge induced by the matter coupling. (The detailed form of $P(\pi^2)$ does not have a particular role in our discussion, since $M_\pi$ depends only on its value at the origin.) 
The coefficients $A$ and $B$ in Eq.~\eqref{eq:RadialEOM} are given by
\begin{align}
    B(\pi,\pi',\lambda,r) &= 2 (1 - H^2 r^2) \llp  r^{2} - 3\lambda \lp H^{2}r\,\pi+(1-H^{2}r^{2})\pi' \rp \lp 3H^{2}r\,\pi+(1-3H^{2}r^{2})\pi' \rp \rrp \,,\label{eq:Bdef}\\
    A(\pi,\pi',\lambda,r) &= 4 \lp 2H^{2}r\,\pi+(1-2H^{2}r^{2})\pi' \rp \llp r-3H^{2}\lambda(\pi-r\pi')\lp H^2r\pi+(1-H^2 r^2)\pi' \rp  \rrp \,.
\label{eq:Adef}
\end{align}

\subsection{Asymptotic solutions: the Vainshtein radius}
\label{sec:analytical}

To gain some intuition for~\eqref{eq:RadialEOM}, we study two asymptotic regimes: near the source at the origin, and near the horizon. Far from the source, the field is expected to be weak and the dynamics governed by the quadratic Lagrangian ${\cal L}_{(2)}$, which determines the large-distance behavior. Close to the source, nonlinearities become important and the quartic interaction ${\cal L}_{(4)}$ dominates. 
The corresponding asymptotic solutions are
\begin{align}
    \pi_{(2)}(r) &= \frac{c_2}{r}(1-3 H^2 r^2) +\frac{d_2}{r}\lp 3 H r+(1-3 H^2 r^2)\,\tanh^{-1} \lp H r \rp\rp \,,\label{eq:pi2solnC2D2}\\ 
    \pi_{(4)}(r) &=  c_4 H + d_4 H^2 r \,,\label{eq:pi4solnC4D4}
\end{align}
where $c_2,d_2,c_4,d_4$ are dimensionless integration constants.

Consider first the case $\lambda=0$, so that there is no quartic term in the Lagrangian and \eqref{eq:pi2solnC2D2} is the full solution.   Integrating the equations of motion across $r=0$ fixes $c_2$ in terms of $M_\pi$: 
\be c_2 = - \frac{M_\pi}{4 \pi}\, .\label{c2sole}\ee  
The other integration constant must be fixed by a separate boundary condition.  A natural choice is to impose regularity at the horizon $r=1/H$;  since $\tanh^{-1} (H r)$ diverges at $r = 1/H$, regularity at the horizon requires 
\be d_2=0\, .\label{d2sole}\ee  
The linear solution is thus
\be  
    \pi_{(2)}=-\frac{M_\pi}{4\pi}\,(1-3 H^2 r^2)\frac{1}{r} \,.\label{linearsol2e}
\ee
This linearized solution already differs qualitatively from its flat-space counterpart~\cite{Andrews:2010km}: it approaches the flat space Coulomb solution $\sim 1/r$ near the origin, but then grows like $\sim r$ as we approach horizon distances.

Now, taking $\lambda\not=0$ we expect the solution near $r=0$ to take the form \eqref{eq:pi4solnC4D4}.
Integrating the equations of motion across $r=0$ fixes $d_4$ in terms of the source mass $M_\pi$ and $\lambda$, and $c_4$ becomes the value of the field at the origin, 
\be\label{eq:c2c4d4}
c_4 = \frac{\pi(0)}{H} \,, \quad d_4 = - \frac{1}{H^2}\lp \frac{M_\pi}{4\pi  \lambda}\rp^{1/3}\,,
\ee 
so that the solution near the origin becomes
\be 
    \pi_{(4)}=\pi(0)-\lp \frac{M_\pi}{4\pi\lambda}\rp^{1/3}r\,.\label{nonlinearsol2e}
\ee
At this stage, the usual strategy is to follow the solution to $r=1/H$, and enforce regularity at the horizon by adjusting $\pi(0)$, which would fix $\pi(0)$, whereupon the full solution would approach the linear solution \eqref{linearsol2e} sufficiently far from the source.  However, we will see below that this is not always possible, and that the solution can become singular before reaching the horizon.

The crossover between the linear behavior \eqref{linearsol2e} and nonlinear behavior \eqref{nonlinearsol2e} defines the Vainshtein radius~\cite{Vainshtein:1972sx}. Equating~\eqref{linearsol2e} and~\eqref{nonlinearsol2e} yields a quadratic equation for $r_{\rm V}$, with solution
\be \label{eq:VainshteinRadius}
    r_{\rm V} = -\frac{c_4 \pm \sqrt{c_4^2 + 4 c_2 \lp  d_4 + 3 c_2\rp}}{ 2 H \lp  d_4 + 3  c_2\rp } 
\ee 
with $\{ c_2, c_4, d_4 \}$ the integration constants in~\eqref{c2sole},~\eqref{eq:c2c4d4}.
Among the two roots in Eq.~\eqref{eq:VainshteinRadius}, the physical Vainshtein radius is the one that is real, positive, and lies within the static patch.

\subsection{Stiffness analysis: the strong-coupling radius}
\label{sec:stiff}
The radial evolution becomes singular whenever $\pi''$ ceases to be determined in a regular way.  Outside the source, $r > 0$, we may write Eq.~\eqref{eq:RadialEOM} as
\be
\pi''=-\frac{A(\pi,\pi',\lambda,r)}{B(\pi,\pi',\lambda,r)}\,,
\ee
which shows that the equation can become stiff in two scenarios: either when $B=0$ while $A \neq 0$, or if $|A|\to\infty$ while $B$ remains finite. In both cases, the effective equation breaks down. We therefore define a {\it strong-coupling radius}, $r_{\rm SC}$, as a point where either 
\begin{align}\label{eq:StiffPoints}
\begin{cases}
    A(r_{\rm SC})  \neq 0  \qquad &\mbox{and} \qquad B(r_{\rm SC}) = 0 \\
    |A(r_{\rm SC})| \to\infty \qquad &\mbox{and} \qquad |B(r_{\rm SC})| < \infty
\end{cases}
\end{align}
The solution cannot in general be extended beyond such a point within the effective description. 
One could hope to solve the equation independently on the two sides of a stiff point and match the resulting solutions across it.  This would, however, require a matching condition, and thus knowledge of the UV physics at play near the stiff point, which we do not have.

To understand when a strong coupling radius appears, we now consider the various cases separately.
\paragraph{Case 1: $A$ and $B$ vanish.}
Evaluating $B$ on the asymptotic solutions~\eqref{eq:pi2solnC2D2} and~\eqref{eq:pi4solnC4D4}, gives
\begin{align}
   \lim_{r\to 1/H^-} B &= \frac{4}{H^2} \lp 1+ 12 \lambda  H^6 c_2^2 \rp (1-H r)\,, \label{eq:Bhorizon}\\
    \lim_{r\to 0^+}  B &= -6 \lambda d_4^2 H^4 \,,\label{eq:Borigin}
\end{align}
with $\{ c_2 , d_2,c_4, d_4 \}$ given in ~\eqref{c2sole},~\eqref{d2sole},~\eqref{eq:c2c4d4}. 
The sign of $B$ in the interval $ r \in (0,1/H)$ depends on the sign and magnitude of $\lambda$. There are three cases: if $\lambda  < - 1  / (12 H^6  c_2^2)  $, then $B_{r\rightarrow 1/H^-}<0$ and $B_{r\to 0}>0$, so continuity forces $B$ to vanish at at least one point in the static patch. If $- 1  / (12 H^6  c_2^2)  < \lambda<0$, then both limits are positive, so a sign change is not forced. If $\lambda>0$, then $B_{r\rightarrow 1/H^-}>0$ and $B_{r\to 0}<0$, so $B$ must again vanish somewhere in the static patch.
This analysis motivates defining a first estimate of the strong-coupling radius from the vanishing of $B$. Using the inner approximate solution~\eqref{eq:c2c4d4} and substituting it into~\eqref{eq:Bdef}, it is then easy to check that $B$ vanishes at
\be\label{eq:StrongCouplingRadius}
    r_{\rm SC} = \frac{6\lambda H^6 c_4 d_4 \pm \sqrt{3  \lambda H^6 d_4^2\lp 1+3\lambda H^6 c_4^2\rp }}{H(1-9\lambda H^6 c_4^2)}\,.
\ee
As in the case of evaluating $r_{\rm V}$, among the two roots in Eq.~\eqref{eq:StrongCouplingRadius} one should pick the one that is real and within the integration domain. Interestingly, we see that $r_{\rm SC}$ is real provided that
\be\label{eq:StrongCouplingCondition}
    \lambda \lp 1+3\lambda H^6 c_4^2\rp \ge 0\,,
\ee
and complex otherwise. This corroborates the previous analysis: for $\lambda >0$, a strong coupling scale is always reached, while there is no real solution for $r_{\rm SC}$ when $ - (3 c_4^2 H^6)^{-1} < \lambda  < 0$.

Next, we analyze whether it is possible for $A$ and $B$ to vanish simultaneously.  To this end, it is useful to introduce
\be
    u(r)\equiv H^2 (\pi-r\pi')\,,\qquad v(r)\equiv H^2 r\pi+(1-H^2 r^2)\pi'\,. \label{eq:uv}
\ee
In terms of these variables,
\begin{align}
    B &= 2(1- H^2 r^2) \llp r^2-3\lambda\,v(v+2 ur) \rrp \label{eq:Buv}\,,\\
    A &= 4( u r+v)\lp r-3 \lambda uv\rp \,.\label{eq:Auv}
\end{align}
Hence $A=0$ implies either
\be
ur+v=0 \qquad \text{or} \qquad r-3\lambda uv=0\,.
\ee
Substituting these conditions back into $B$ yields
\begin{align}
    B\Big|_{ur+v=0} &= 2(1-H^2 r^2)\lp r^2+3\lambda v^2\rp \,,\\
    B\Big|_{r-3\lambda uv=0} &= -2(1-H^2 r^2)\lp r^2+3\lambda v^2\rp \,.
\end{align}
For $\lambda>0$, the factor in parentheses is strictly positive for real $v$ and $r \in (0,1/H)$. It follows that there are no points inside the static patch at which $A=B=0$. Therefore, whenever $B$ crosses zero for $\lambda>0$, the equation is genuinely stiff and the radial solution cannot be continued smoothly through that point. By contrast, for $\lambda<0$ the condition $A=B=0$ may in principle be satisfied, so regular solutions are not excluded on these grounds.

\paragraph{Case 2: $A$ and $B$ diverge.}

Unlike the condition $B=0$, which is algebraic and can be studied directly on the asymptotic branches, the condition $|A|\to\infty$ depends on the actual behavior of the full solution $\pi(r)$ and its derivative. For instance, evaluating $A$ on the outer approximate solution gives
\be
A(\pi_{(2)},\pi'_{(2)},\lambda,r\ll 1)\sim -\frac{c^3_2 \lambda H^2}{r^5}\,,
\ee
so that, if one integrates inward from the horizon and the linear branch persists too far toward the origin, $A$ diverges. In other words, the quartic branch must dominate sufficiently early; otherwise the system becomes stiff before the source is reached.

\subsection{Screening in de Sitter space}
\label{sec:vainshtein}
We have seen that the solutions are controlled by two scales: the {\it Vainshtein radius} $r_{\rm V}$ of Eq.~\eqref{eq:VainshteinRadius}, where the solution transitions between nonlinear and linear regimes, and the {\it strong-coupling radius} $r_{\rm SC}$ of Eq.~\eqref{eq:StrongCouplingRadius}, where the radial equation becomes singular.  We call a screened profile one in which there are clear linear and non-linear regimes, separated by a Vainshtein radius, before any strong coupling radius is reached.  The existence of a screening solution thus depends on the relative ordering between the Vainshtein and strong coupling radius.    We will see that the existence and ordering of these radii also depend on how the solution is constructed, namely, whether we attempt to fix boundary conditions at the horizon and integrate inwards, or we fix boundary conditions at the origin and integrate outwards.

If the radial equation~\eqref{eq:RadialEOM} is integrated outward from the origin, then  in order to have a screening solution the screened branch must reach the linear regime before encountering the singularity, requiring
\be
    0<r_{\rm V}<r_{\rm SC}< \frac{1}{H}\,.
\ee
If instead the equation is integrated inward from the horizon, the linear branch must cross over to the non-linear solution before the strong coupling radius is reached, implying
\be
    0<r_{\rm SC}<r_{\rm V}<\frac{1}{H}\,.
\ee
Thus, unlike in flat space, the existence of a Vainshtein radius alone does not guarantee a viable screening solution: it must also be appropriately ordered relative to the strong-coupling scale, if one exists.

Using Eqs.~\eqref{eq:StrongCouplingRadius} and~\eqref{eq:VainshteinRadius}, we can gain an understanding of the relative order of magnitude of these two scales. Using Eq.~\eqref{eq:c2c4d4}, we consider the case where $|\lambda| H^6 c_4^2 \ll 1$ and $|c_4| \ll  |M_\pi|$. 
Then the strong-coupling and Vainshtein radii are estimated by 
\be 
    r_{\rm SC} \sim |M_\pi|^\frac13 \lambda^\frac16\,, \qquad r_{\rm V} \sim \lp  M_\pi^{-\frac23} \lambda^{-\frac13} + 3 H^2 \rp^{-\frac12}\,,
\ee 
showing that 
\be 
    r_{\rm V} \sim \frac{ r_{\rm SC}}{ \sqrt{1 + 3 H^2  r^2_{\rm SC}}}    \leq  r_{\rm SC}\,.
\ee
This demonstrates that the solution can transition from the non-linear to the linear regime before encountering the strong-coupling scale, while the two radii become of the same order of magnitude $r_{\rm V} \sim r_{\rm SC}$ when the strong-coupling scale is well sub-horizon, i.e. $r_{\rm SC} \ll 1/H$. 
Interestingly, $|M_\pi|^\frac13 \lambda^\frac16$ matches the Vainshtein scale of the flat-space problem~\cite{Andrews:2010km}. Thus, curvature pushes $r_{\rm V}$ inward, while producing a strong-coupling radius \textit{at} the flat-space counterpart Vainshtein radius. 
In the case $ |M_\pi |\gg | c_4 |\gg 1/|\lambda H^6|$, the asymptotic behaviors of the two scales are 
\be 
    r_{\rm SC} \sim \frac{|M_\pi|^\frac13 ( \lambda H^6)^{-\frac13} }{H c_4}\,, \qquad r_{\rm V} \sim \frac{1}{\sqrt{3} H}\,,
\ee 
and the order of magnitude and relative order between $r_{\rm SC}$ and $r_{\rm V}$ depend entirely on the choice of parameters.  
As a final note, we emphasize that the value of $r_{SC}$ obtained from Eq.~\eqref{eq:StrongCouplingRadius} is approximate because it is found using the approximate solution~\eqref{eq:pi4solnC4D4}, which is only accurate close to the origin. While the exact value of the strong-coupling scale should be found numerically, this analytic estimate can still provide a useful order of magnitude estimate for comparison with the Vainshtein radius $r_V$.

\section{Numerical solutions}
\label{sec:numerical}

To see how this all plays out, we now study Eq.~\eqref{eq:RadialEOM} numerically. Our aim is to understand how the fully nonlinear solutions behave as we attempt to evolve between the origin and the horizon, and to determine whether this evolution breaks down before the entire static patch is covered.

\subsection{Dimensionless variables}
We will perform the numerical integration in dimensionless quantities in units of the Hubble scale $H$. Thus, we perform the rescaling
\be \label{eq:rescaling}
    \pi \to H \pi \,, \qquad r \to \frac{r}{H} \,, \qquad \lambda \to  H^{-6} \lambda\,,
\ee 
so that $ r \in (0,1)$. The radial equation in these coordinates can be obtained from Eq.~\eqref{eq:RadialEOM} simply by setting $H=1$.  We also rescale the mass of the source, choosing
\be 
M \to H M\,,
\ee 
which enters the equation through $M_\pi$, so that $M_\pi$ doesn't change.
We stress that we use these dimensionless quantities only when presenting numerical results, hence in Secs.~\ref{sec:numerical1} and~\ref{sec:numerical2}, and later in Sec.~\ref{sec:NumericsPerts}.

\subsection{Scaling relations}\label{sec:numerical1}
Before presenting explicit solutions, it is useful to first note that for $\lambda\neq 0$ the radial equation enjoys a scaling symmetry: one can verify that
\begin{align}
\label{eq:ABLambdaTransf}
B\lp \frac{\pi}{\sqrt{|\lambda|}}, \frac{\pi'}{\sqrt{|\lambda|}}, \lambda, r \rp &= B\lp \pi,\pi', {\rm sign}(\lambda), r \rp \,, \\
A\lp \frac{\pi}{\sqrt{|\lambda|}}, \frac{\pi'}{\sqrt{|\lambda|}}, \lambda, r \rp &= \frac{A\lp \pi,\pi', {\rm sign}(\lambda), r \rp}{\sqrt{|\lambda|}}\,,
\end{align}
where ${\rm sign}(\lambda)=\pm 1$ according to whether $ \lambda$ is positive or negative. Therefore, if $\pi_{{\rm sign}(\lambda)}(r)$ denotes a solution of Eq.~\eqref{eq:RadialEOM} for $\lambda=\pm 1$, then the corresponding solution for generic nonzero $\lambda$ is obtained as
\be
\pi_{\lambda}(r)=\frac{\pi_{{\rm sign}(\lambda)}(r)}{\sqrt{|\lambda|}}\,,
\ee
with boundary data scaled accordingly
\be
\pi_{\lambda}(0)=\frac{\pi_{0}^{{\rm sign}(\lambda)}}{\sqrt{|\lambda|}}\,,
\qquad
\pi'_{\lambda}(0)=\frac{{\pi'_{0}}^{{\rm sign}(\lambda)}}{\sqrt{|\lambda|}}\,,
\ee
with  $\pi_{0}^{{\rm sign}(\lambda)}$ and ${\pi'_{0}}^{{\rm sign}(\lambda)}$  the values of the $\pi_{{\rm sign}(\lambda)}(r)$ and its derivative at the origin. It is therefore sufficient to study only the two representative cases $\lambda=\pm 1$. (The case $\lambda=0$ is where the quartic interaction is absent and the solution in Eq.~\eqref{eq:pi2solnC2D2} can be extended to the whole integration domain.)
This scaling property also shows that the location of the strong coupling scale \eqref{eq:StiffPoints} depends only on the sign of $\lambda$, not on its magnitude, since the presence of the strong-coupling radius associated with the conditions $B=0$ or $A \to \infty$ is invariant under the above rescaling. 

\subsection{Choice of boundary conditions}\label{sec:numerical2}
There are two natural ways to specify boundary data for the radial equation---integrating inward from the horizon or outward from the origin---and we will see that these lead to different types of solutions.  
In the first case, it is natural to impose regularity at the horizon as a boundary condition, i.e. use Eq.~\eqref{eq:pi2solnC2D2} to remove the divergent $\operatorname{tanh}^{-1}$ branch and select a solution that is regular as $r \to 1$.  On the other hand, if we are integrating outwards from the origin, we instead specify boundary data there, fixing $\pi(0)$ and $M_\pi$.  
Note that it is not possible to know a priori which choice of these two constants guarantees that the solution will be regular at the horizon. However, if the evolution encounters a strong-coupling radius, where the equation becomes singular and the description breaks down before reaching the horizon, then any formal continuation to $r=1$ becomes physically irrelevant.  In other words: horizon regularity is a meaningful criterion for selecting physical solutions only if the solutions themselves can be extended within the whole static patch. Thus, the presence or absence of a strong-coupling radius determines whether imposing regularity at the horizon is necessary.

\subsubsection{Integrating outward from the origin}
\label{sect:intfromorigin}
\begin{figure}[H]
    \centering
    \includegraphics[width=0.48\linewidth]{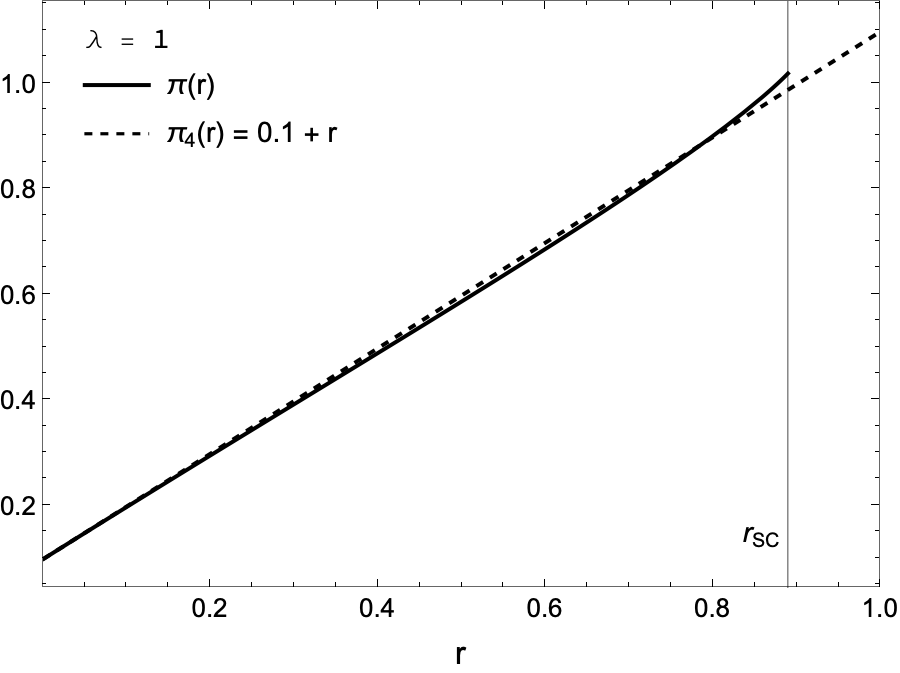}
    \includegraphics[width=0.48\linewidth]{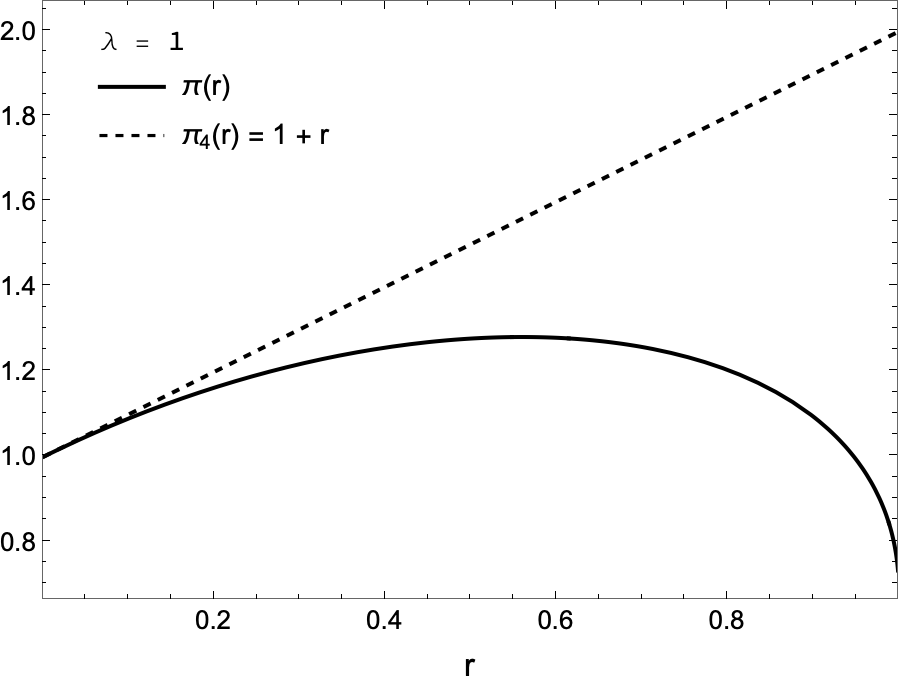}
    \includegraphics[width=0.48\linewidth]{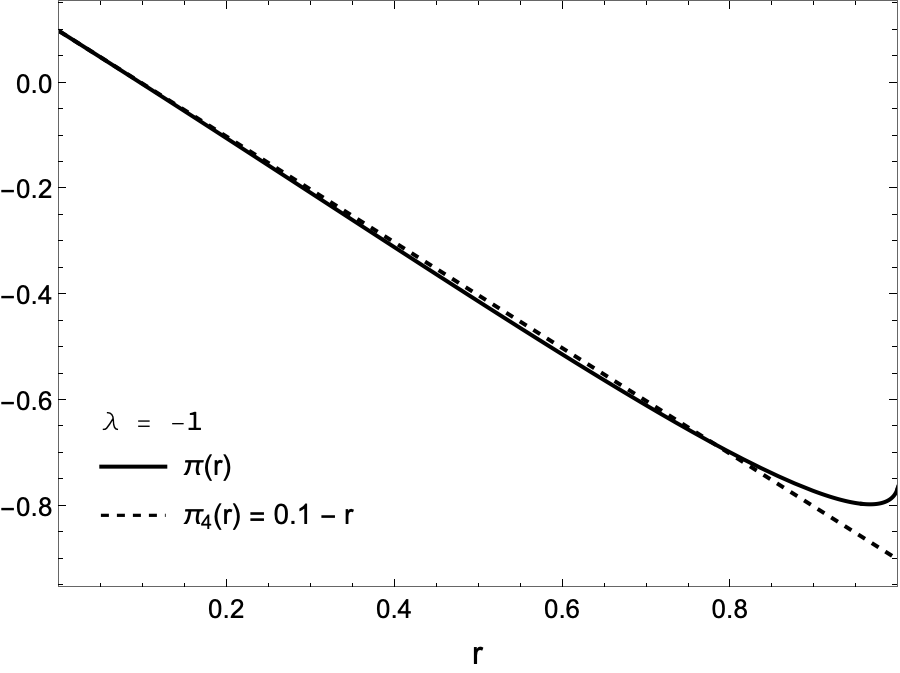}
    \includegraphics[width=0.48\linewidth]{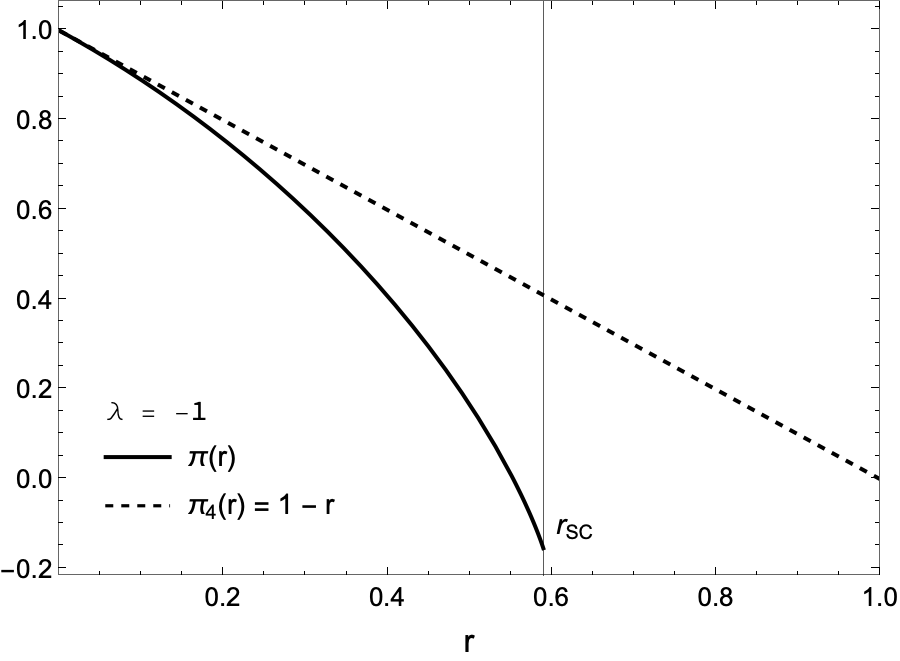}
    \caption{Numerical solutions obtained by integrating from the origin, for $\lambda= \pm 1$ (top and bottom row respectively). Dashed lines denote analytic approximations and solid lines the corresponding numerical solutions. In each panel, boundary conditions are chosen to match the analytic profile of the near-origin solution; for example, in the top-left panel $\pi(0)=0.1$ and $\pi'(0)=1$. The vertical line denotes the $r_{SC}$ scale, if one is present inside the horizon. }
    \label{fig:Plot12}
\end{figure}

Fig.~\ref{fig:Plot12} shows numerical solutions obtained by integrating outward from the origin for $c_4=0.1$ and $c_4=1$. 
For $\lambda=1$ (top row), we see that the strong coupling radius is inside the static patch for $c_4=0.1$, while it is pushed beyond the horizon for $c_4=1$. In the latter case, the solution also deviates significantly from the linear profile $\pi_4(r)$, indicating the onset of screening. 
One can estimate the value of $c_4$ such that $r_{\rm SC}  = 1$, so that the strong-coupling scale reaches the boundary of the static patch. Using $\lambda = d_4 = 1$, as in the first row of Fig.~\ref{fig:Plot12}, and Eq.~\eqref{eq:StrongCouplingRadius} one finds 
\be 
 r_{\rm SC} = 1 \qquad \to \qquad c_4 \approx 0.39 \,,
\ee 
which is supported numerically  by Fig.~\ref{fig:Plot12} because $c_4$ is indeed between the values $c_4 = 0.1$ (left panel) and $c_4 = 1$ (right panel).
For $\lambda=-1$ (bottom row), a strong-coupling radius exists only if $c_4^2>1/3$ (satisfied in the bottom-right panel). The two panels illustrate the corresponding cases in which this scale is absent or present within the static patch. 
In all the panels, we have fixed $M_\pi = - 4 \pi$, as can be understood from Eq.~\eqref{nonlinearsol2e}, considering that we have picked $d_4 = 1$ in the top row, and $d_4 = -1$ in the bottom one.

\subsubsection{Integrating from the horizon}
\begin{figure}[h]
    \centering
    \includegraphics[width=0.48\linewidth]{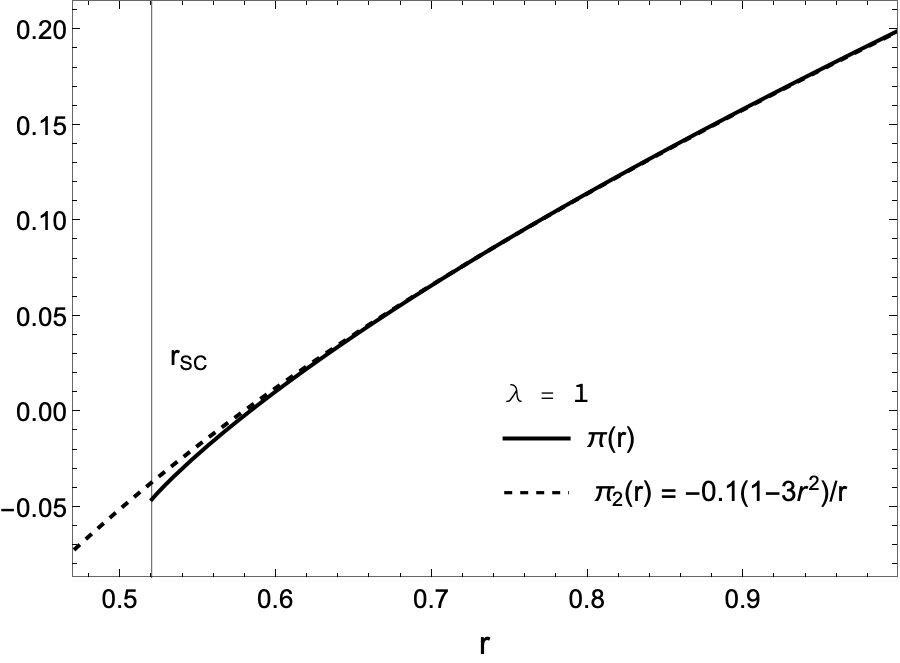}
    \includegraphics[width=0.48\linewidth]{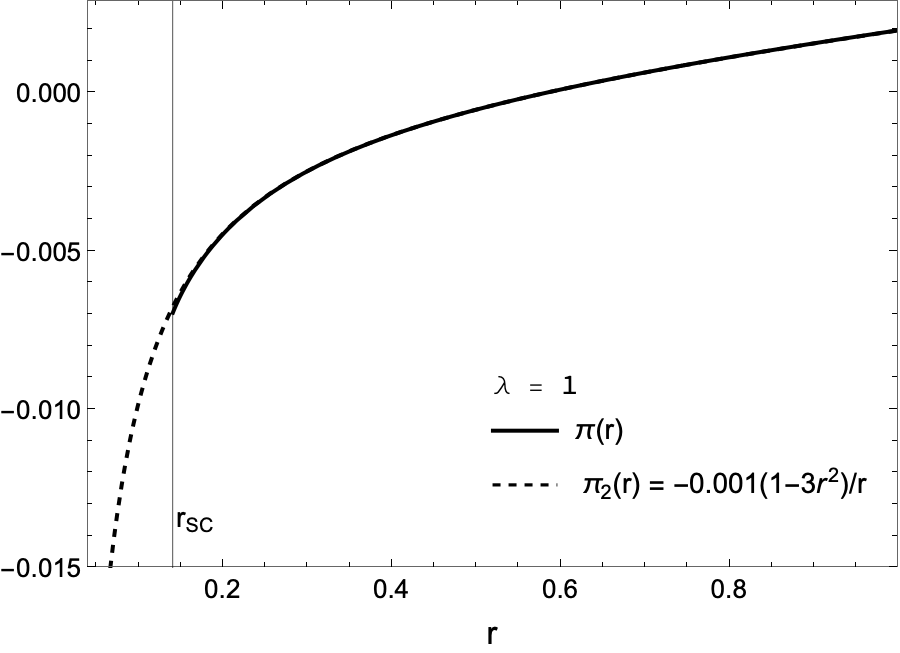}
    \includegraphics[width=0.48\linewidth]{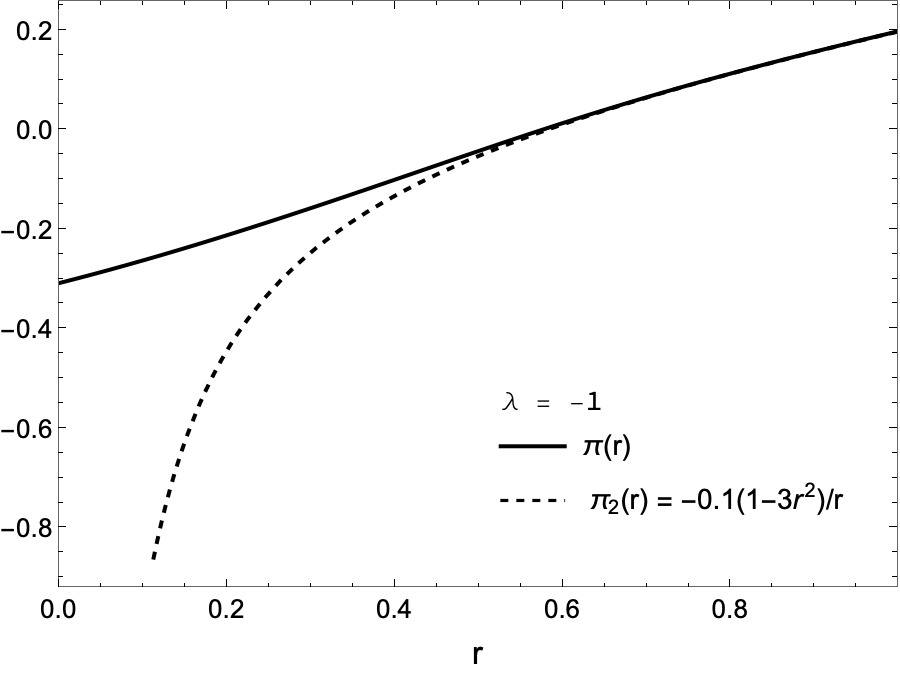}
    \includegraphics[width=0.48\linewidth]{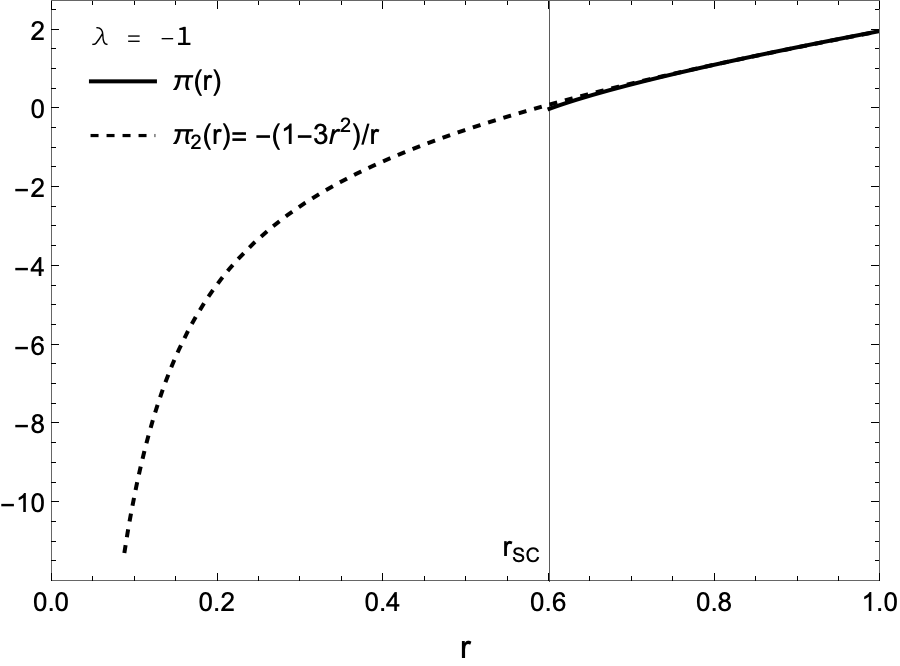}
    \caption{Numerical solutions obtained by integrating from the horizon, with $\lambda= \pm 1$ in the top and bottom row respectively. Dashed lines denote the analytic profile $\pi_{(2)} = c_2 (1 - 3 r^2)/r$, with $c_2$ given in the legend, and solid lines the corresponding numerical solutions. Boundary conditions are imposed at the horizon to match this profile, $\pi(1)=-2c_2$ and $\pi'(1)=-4c_2$.}
    \label{fig:Plot34}
\end{figure}

Fig.~\ref{fig:Plot34} shows solutions obtained by integrating inward from the horizon.  Here, boundary conditions are set by matching to the linear solution near the horizon, where regularity at the horizon is imposed by taking $d_2=0$.  
The key criterion is whether $B$ changes sign within the static patch. From Eqs.~\eqref{eq:Borigin} and~\eqref{eq:Bhorizon}, for $\lambda>0$ this is unavoidable: $B$ must vanish for some $0<r<1$, implying the existence of a strong-coupling radius. The first row illustrates this, with the location of the zero controlled by the choice of $c_2 = \{ -0.1, -0.001\}$ in the two panels.
For $\lambda=-1$, the sign of $B$ near the horizon depends on $1-12c_2^2$. A strong-coupling point arises only if this quantity is negative. The second row shows both cases, with $c_2 = -0.1$ in the left panel, and $c_2 = -1$ in the right panel. 
The numerical solutions match well their analytical expectation at large scales (solid vs dashed lines).

\section{Perturbation analysis}
\label{sec:perturbation}

In the previous section, we found a static, spherically symmetric  solution of the equations of motion. Here we assess their stability by studying perturbations around these backgrounds. We expand the scalar fields as
\be
\pi^I(t,r,\theta,\varphi)=\pi^I_0(r)+\phi^I(t,r,\theta,\varphi)\,,
\ee
where $\pi^I_0(r)=\delta^I_N\,\pi_0(r)$ and $\pi_0(r)$ is the  radial solution found in Section~\ref{sec:radial}. 
The full quadratic action is derived in 
Appendix~\ref{app:CoefficientQuadraticAction}, and takes the form
\begin{align}
    \frac{1}{\sqrt{-g}}\,{\cal L}^{\rm tot}_{\rm quad}&= -{\cal K}^{\mu\nu}_N\,\partial_\mu\phi^N\partial_\nu\phi_N + {\cal M}_N\,\phi^N\phi_N -{\cal K}^{\mu\nu}_{\hat I}\,\partial_\mu\phi^{\hat I} \partial_\nu\phi_{\hat I} +{\cal M}_{\hat I}\,\phi^{\hat I}\phi_{\hat I}\,,
\label{eq:QuadLagrangianAligned}
\end{align}
where $\hat{I}=1,\ldots,N-1$.
The coefficients $\{{\cal M}_N, {\cal M}_{\hat I} \}$ of the mass-like terms are given by
\begin{align}
    {\cal M}_N&= 
     \frac{4 H^2(1 - H^2 r^2 )}{B r^2 } \lp 2 r^4 + 3 r^2 (v - u r)^2 \lambda + 9\lambda^2  v^2 (v^2 + 2 u v r + 3 u^2 r^2) \rp\,,\label{eq:MassNtermnoH} \\
    {\cal M}_{\hat I}&=  \frac{8 H^2 + {\cal M}_N}{3}\,, \label{eq:MassItermnoH}
\end{align}
where $u,v$ are the variables introduced in Eq.~\eqref{eq:uv}, and $B$ is the coefficient of the second radial derivative of Eq.~\eqref{eq:Buv}. 
The tensors $\{ {\cal K}^{\mu\nu}_N ,{\cal K}^{\mu\nu}_{\hat I}\}$ in front of the derivative terms are diagonal and given by,
\begin{align}
    {\cal K}^{\mu\nu}_N&={\rm diag}\lp  g^{tt}{\cal K}^t_N,\, g^{rr}{\cal K}^r_N,\, g^{\theta\theta}{\cal K}^\theta_N,\, g^{\varphi\varphi}{\cal K}^\theta_N \rp \,, \label{eq:KineticNterm}\\
    {\cal K}^{\mu\nu}_{\hat I}&={\rm diag}\lp  g^{tt}{\cal K}^t_{\hat I},\,g^{rr}{\cal K}^r_{\hat I},\,g^{\theta\theta}{\cal K}^\theta_{\hat I},\,g^{\varphi\varphi}{\cal K}^\theta_{\hat I}\rp \,, \label{eq:KineticIterm}
\end{align}
with $\{ {\cal K}^t_N,{\cal K}^r_N,{\cal K}^\theta_N \}$ and  $\{ {\cal K}^t_{\hat I},{\cal K}^r_{\hat I},{\cal K}^\theta_{\hat I} \}$ functions of $\pi_0, \pi'_0, \pi''_0$ and $r$, which we provide explicitly in Appendix~\ref{app:CoefficientQuadraticAction}.

For a fully stable solution, all of these coefficients should be positive.  A wrong-sign kinetic term, ${\cal K}^t<0$, signals a ghost-like instability.  A wrong-sign   for the spatial derivative coefficients, ${\cal K}^r<0$ or ${\cal K}^\theta<0$, indicates a gradient instability \cite{Joyce:2014kja}.  A wrong sign for the mass-like terms, ${\cal M}<0$, signals a tachyonic instability.

The  radial and angular propagation speeds are defined from the ratios of the kinetic coefficients,
\be\label{eq:PropSpeed}
    (c^r)^2 \equiv \frac{{\cal K}^r}{{\cal K}^t} \,,\qquad (c^\theta)^2 \equiv \frac{{\cal K}^\theta}{{\cal K}^t} \,.
\ee
Hence, a wrong relative sign between the spatial and time coefficients of the kinetic terms leads to imaginary propagation speed, and if $|{\cal K}^t|< |{\cal K}^r|, |{\cal K}^\theta|$ then propagation can be superluminal.

\subsection{Analytical considerations}
Before delving into the numerics, it is useful to gain some analytical understanding of the perturbation equations.
It is convenient to rewrite the kinetic coefficients in terms of the variables $u$ and $v$ already introduced in Eq.~\eqref{eq:uv}, and to use the background equation $\pi''=-A/B$ outside the source. For the mode aligned with the background, we obtain
\begin{align}
    {\cal K}^t_N &= \frac{2(1-H^2 r^2)}{r^2}\,\frac{\bigl(r^2+3\lambda v^2\bigr)^2}{B}\,, \label{eq:KNT}\\
    {\cal K}^r_N &= \frac{B}{2r^2(1-H^2 r^2)}\,, \label{eq:KNR}\\
    {\cal K}^\theta_N &= 2(1-H^2 r^2)\,\frac{(1+3\lambda u^2)(r^2+3\lambda v^2)}{B} \label{eq:KNTheta}\,,
\end{align}
These expressions show that the same quantity $B$, which governs the stiffness of the background equation, also controls the kinetic structure of the aligned perturbation. Inside the static patch, the signs of $\{ {\cal K}^t_N, {\cal K}^r_N \}$ are entirely fixed by the sign of $B$, while for ${\cal K}^\theta_N$, positivity requires that $B$ and $(1+3\lambda u^2)(r^2+3\lambda v^2)$ have the same sign.
The corresponding sound speeds are
\be 
    (c_N^r)^2 = \frac{\lp r^2 - 3 \lambda v (v + 2 ur)\rp^2 }{ \lp r^2+3\lambda v^2 \rp^2}\,,  \qquad  (c_N^\theta)^2 = \frac{(1+3\lambda u^2)r^2}{r^2+3\lambda v^2} \,,
\ee 
which must satisfy $ 0 < (c_N^r)^2 , (c_N^\theta)^2 \leq 1$ in order to have stable perturbations that do not propagate superluminally. 
These conditions can be studied analytically on the two asymptotic branches introduced in Eqs.~\eqref{eq:pi2solnC2D2} and~\eqref{eq:pi4solnC4D4}. On the horizon-regular linear branch one finds
\begin{align}\label{eq:KineticCloseHorizon}
    \lim_{r\to 1/H^-} \lp {\cal K}^t_N,\,{\cal K}^r_N,\,{\cal K}^\theta_N \rp = 1+12\lambda H^6 c_2^2 \,,
\end{align}
showing that close to the horizon the sound speeds become luminal on the regular branch, while on the screened near-source branch, the leading asymptotics are instead
\be \label{eq:KineticCloseOrigin}
    \lim_{r\to 0^+} \lp {\cal K}^t_N,\,{\cal K}^r_N \rp = -\frac{3\lambda d^2_4 H^4 }{r^2} \,, \qquad \lim_{r\to 0^+} \lp {\cal K}^\theta_N \rp = - (1 + 3 \lambda H^6 c_4^2) \,,
\ee 
so that the radial mode becomes luminal, while the angular mode has $(c_N^\theta)^2  \to  0$.

\subsubsection{Stability}
For $\lambda>0$, the signs of $\{{\cal K}^t_N, {\cal K}^r_N, {\cal K}^\theta_N\}$ are fixed by the sign of $B$, since the numerator of Eq.~\eqref{eq:KNTheta} does not vanish within the static patch. Perturbations are therefore stable for $B>0$ and unstable otherwise. 
From Sec.~\ref{sec:stiff}, we know that $B$ necessarily changes sign between the origin and the horizon in this case, implying the presence of a strong-coupling point. Whether this point lies within the static patch depends on the boundary conditions.
If $r_{\rm SC}\geq 1/H$, then $B<0$ throughout $0<r<1/H$, and perturbations are unstable everywhere. If $0<r_{\rm SC}<1/H$, the static patch splits into two regions; perturbations are unstable for $r<r_{\rm SC}$ and stable for $r>r_{\rm SC}$.

\bigskip
For $\lambda<0$, stability requires both $B>0$ and $(1+3\lambda u^2)(r^2+3\lambda v^2)>0$. 
From Eqs.~\eqref{eq:Borigin} and~\eqref{eq:Bhorizon}, we know that $B$ is positive close to the origin irrespective of the boundary condition, while close to the horizon its positivity is contingent on $1 - 12 c_2^2 > 0$ (see Eq.~\eqref{eq:KineticCloseHorizon}). Near the origin, Eq.~\eqref{eq:KineticCloseOrigin} shows that $\{{\cal K}^t_N,\,{\cal K}^r_N  \}$ are positive, while ${\cal K}^\theta_N > 0$ only if
\be
|\lambda| H^6 c_4^2> \frac{1}{3}\,.
\ee
Interestingly, this condition is related to the presence of a strong-coupling radius, as shown in Eq.~\eqref{eq:StrongCouplingCondition}.   
This implies that, in the dS spacetime, it is theoretically possible to find stable radial solutions in the range $r \in (0, r_{\rm SC})$ by selecting $\lambda = -1$ and $c_4 > 1/3$. Conversely, solutions that lack a strong coupling scale within the horizon can never be stable.
However, we emphasize that this condition is necessary but not sufficient. The complete behavior of the solution depends on the specific values of $c_4$ and $d_4$, and whether these values lead to the stable branch at the horizon.

\subsubsection{The ${\hat I}$ modes}
While this discussion focused on the $N$ mode,  we point out that it can be straightforwardly extended to the $\hat I$ modes. Indeed, their mass and kinetic coefficients are related by
\be 
 {\cal K}^t_{\hat I} = \frac{2+{\cal K}^t_N}{3} \,, \qquad {\cal K}^r_{\hat I} = \frac{2+{\cal K}^r_N}{3} \,, \qquad {\cal K}^\theta_{\hat I} = \frac{2+{\cal K}^\theta_N}{3}\,, \label{eq:ImodesasNmodes}
\ee 
as shown in Appendix~\ref{app:CoefficientQuadraticAction}, while the mass is given in Eq.~\eqref{eq:MassItermnoH}.
Hence, the transverse sector is completely determined once the aligned mode is known, and one can perform a similar stability analysis.

\subsection{Numerical analysis}\label{sec:NumericsPerts}

We now proceed to numerically study the stability and superluminality of the full solutions.
As before, we perform the numerical computation in dimensionless  variables, performing the rescalings in Eq.~\eqref{eq:rescaling}.

\subsubsection{Perturbation coefficients for $\lambda = 1$}
\begin{figure}
    \centering
    \includegraphics[width=0.48\linewidth]{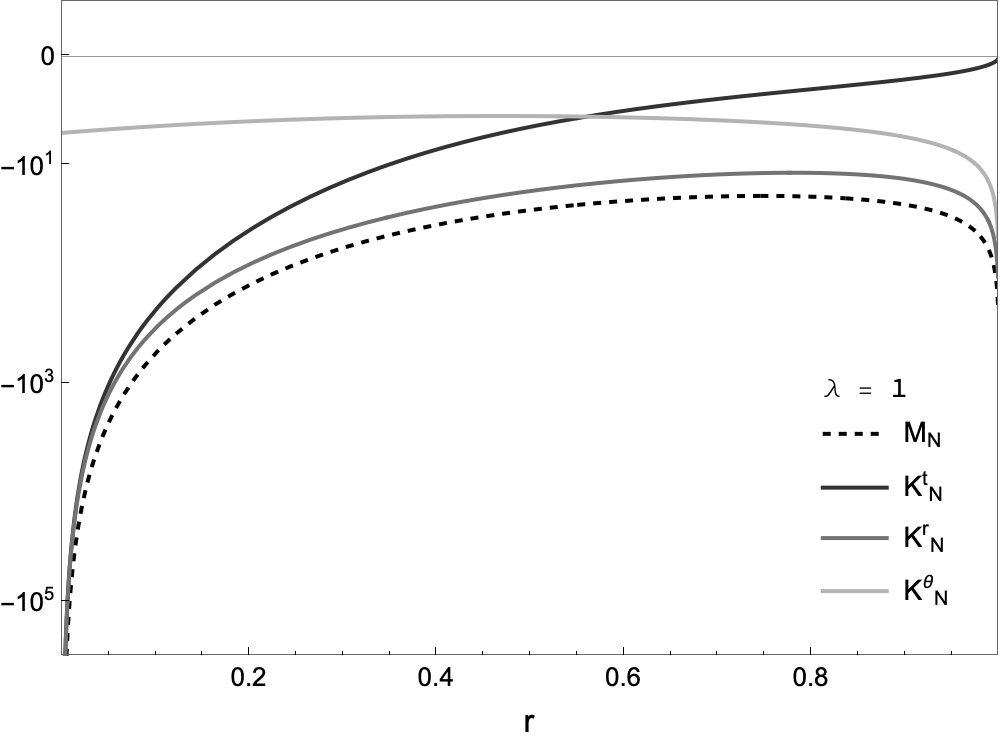}
    \includegraphics[width=0.48\linewidth]{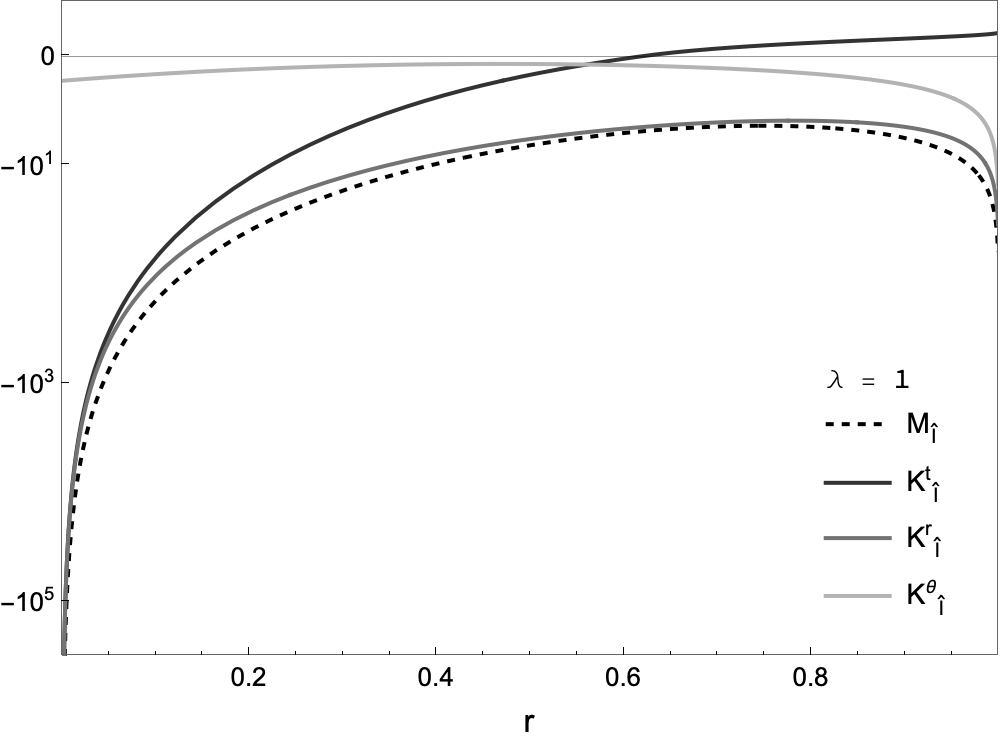}

    \includegraphics[width=0.48\linewidth]{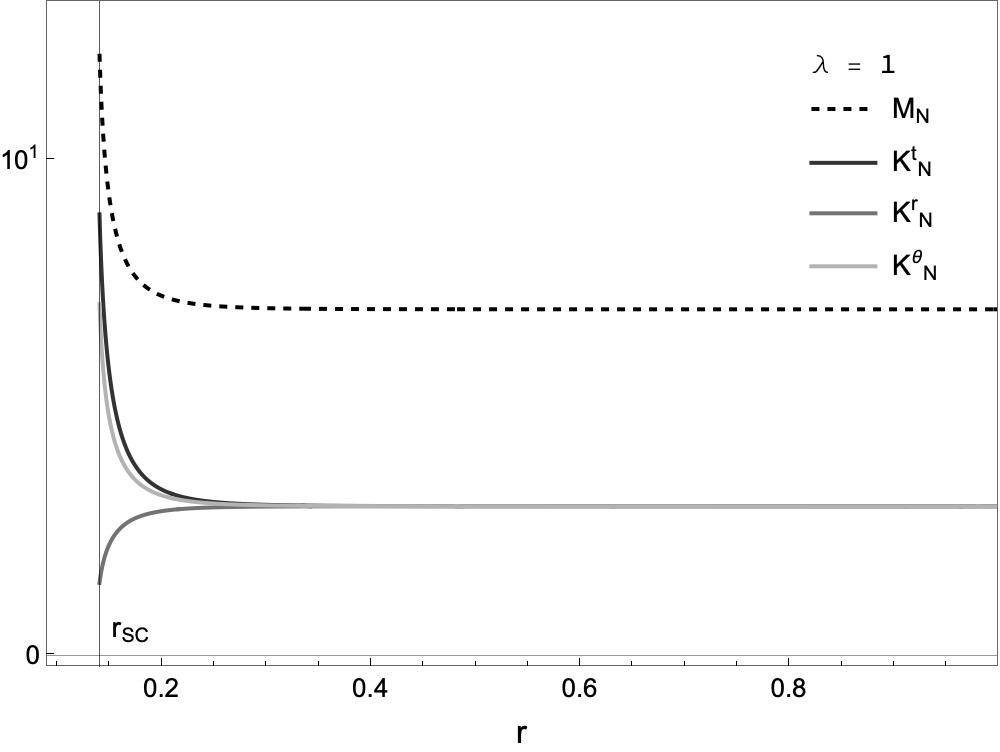}
    \includegraphics[width=0.48\linewidth]{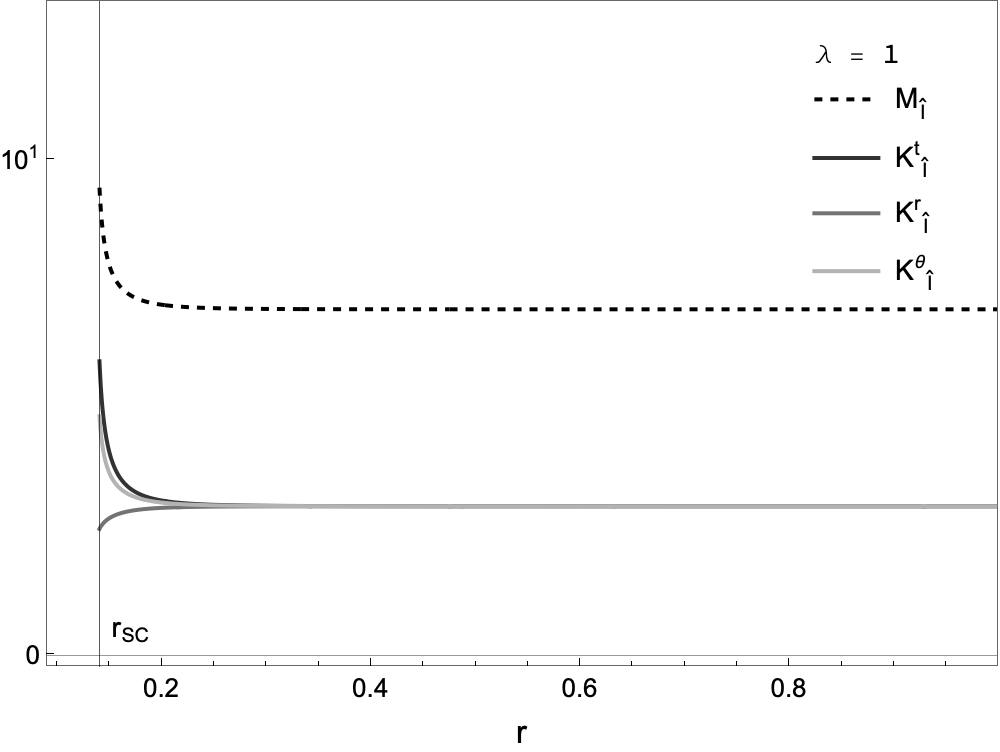}
    \caption{Stability coefficients for $\lambda=1$. The top row shows the solution integrated from the origin (top-right panel of Fig.~\ref{fig:Plot12}), defined over the full static patch. The bottom row shows the solution integrated from the horizon (top-right panel of Fig.~\ref{fig:Plot34}), which contains a strong-coupling scale. The first and second columns correspond to the $N$ and $\hat I$ sectors.}
    \label{fig:StabilityLambda1}
\end{figure}

Fig.~\ref{fig:StabilityLambda1} shows the masses and kinetic coefficients of all perturbation modes for $\lambda=1$. In this case, stability of the $N$ modes is entirely set by the sign of $B$, with $B<0$ for $r\in(0,r_{\rm SC})$ and $B>0$ for $r\in(r_{\rm SC},1)$. If the strong-coupling radius lies outside the horizon, $r_{\rm SC}\geq1$, then $B<0$ throughout the whole static patch. The first column illustrates this through the $N$-mode kinetic coefficients.
The top row corresponds to the solution integrated from the origin with $c_4=1$, $d_4=1$ (top-right panel of Fig.~\ref{fig:Plot12}), for which $r_{\rm SC}>1$ and all coefficients are negative across the static patch. The bottom row uses the solution integrated from the horizon with $c_2=-0.0001$ (top-right panel of Fig.~\ref{fig:Plot34}); here the solution exists only for $r>r_{\rm SC}$, where all coefficients are positive, thus perturbations are stable, and the propagation speeds are subluminal at small radii, and become luminal toward the horizon.
The second column shows the $\hat I$ sector for the same backgrounds, with analogous behavior shifted according to Eq.~\eqref{eq:ImodesasNmodes}.
Thus, we conclude that the numerical solution obtained by integrating from the horizon with boundary conditions $c_2 = -0.001$ and $d_2 = 0$, and with $\lambda = 1$ is stable and perturbations are not superluminal for $r \in (r_{\rm SC}, 1)$.

\subsubsection{Stability for $\lambda = -1$, integrating from the origin}
\begin{figure}
    \centering
    \includegraphics[width=0.48\linewidth]{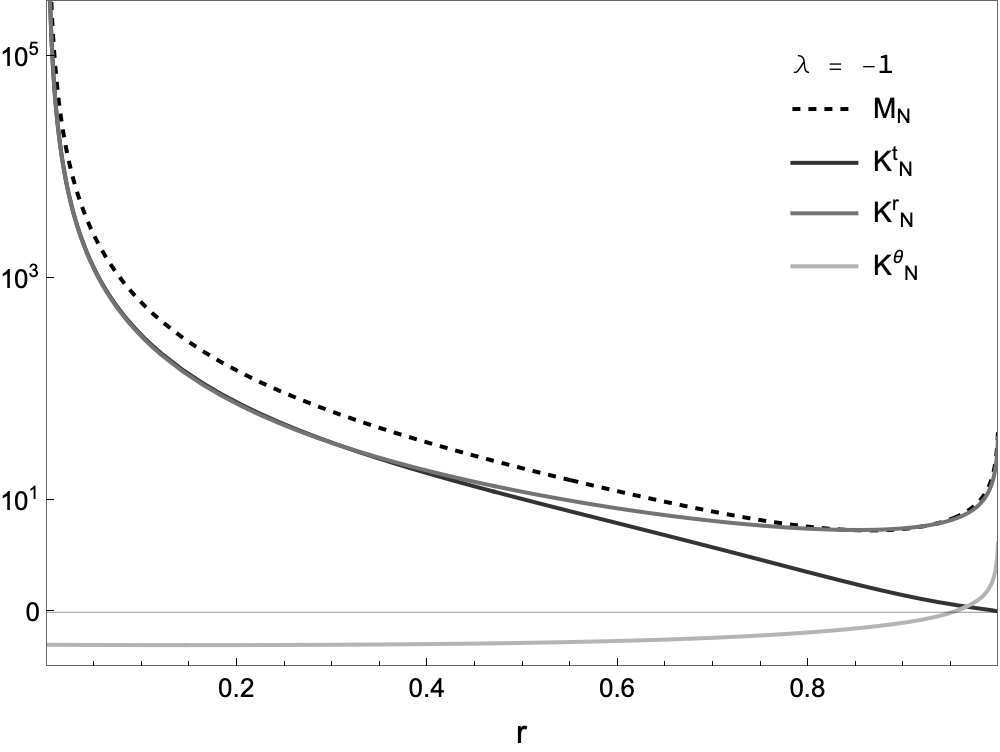}
    \includegraphics[width=0.48\linewidth]{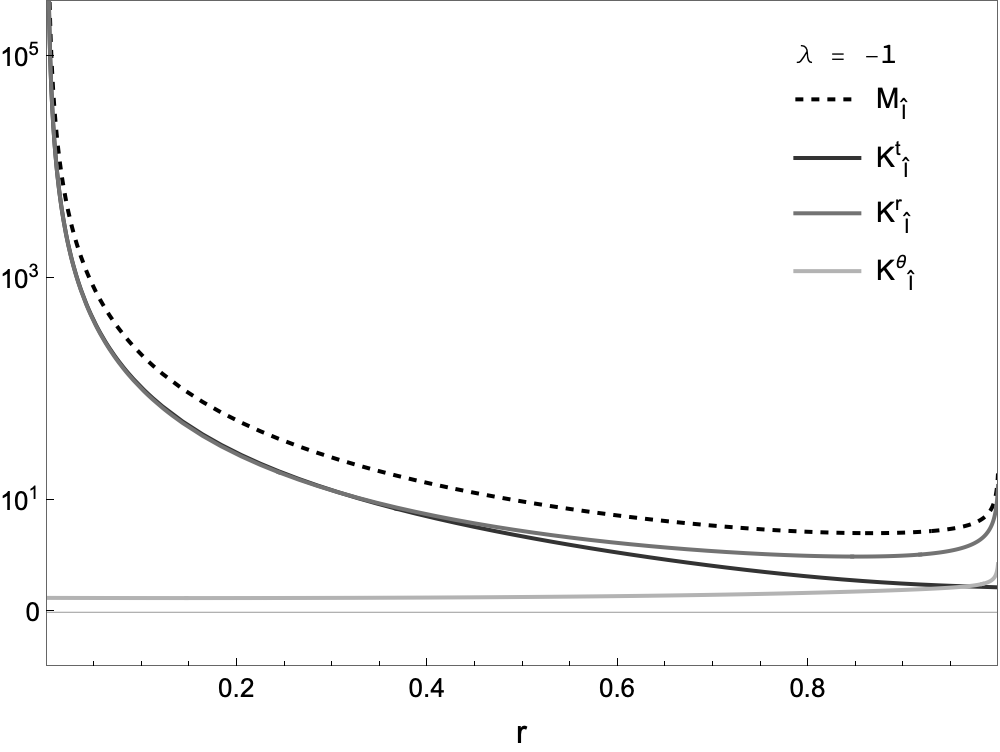}
    
    \includegraphics[width=0.48\linewidth]{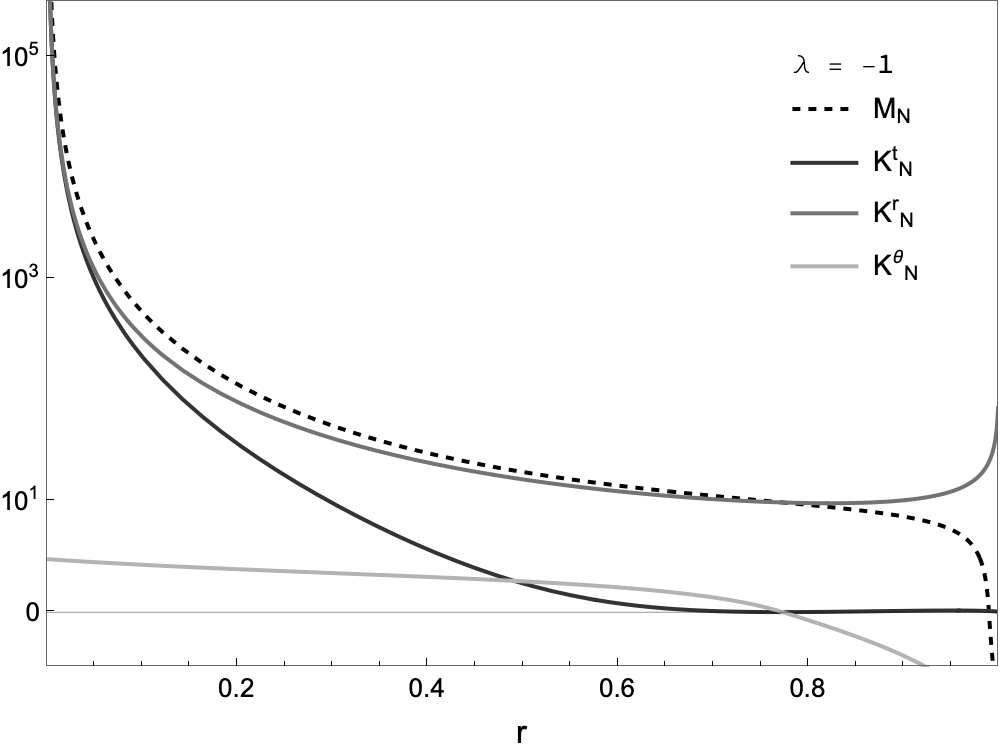}
    \includegraphics[width=0.48\linewidth]{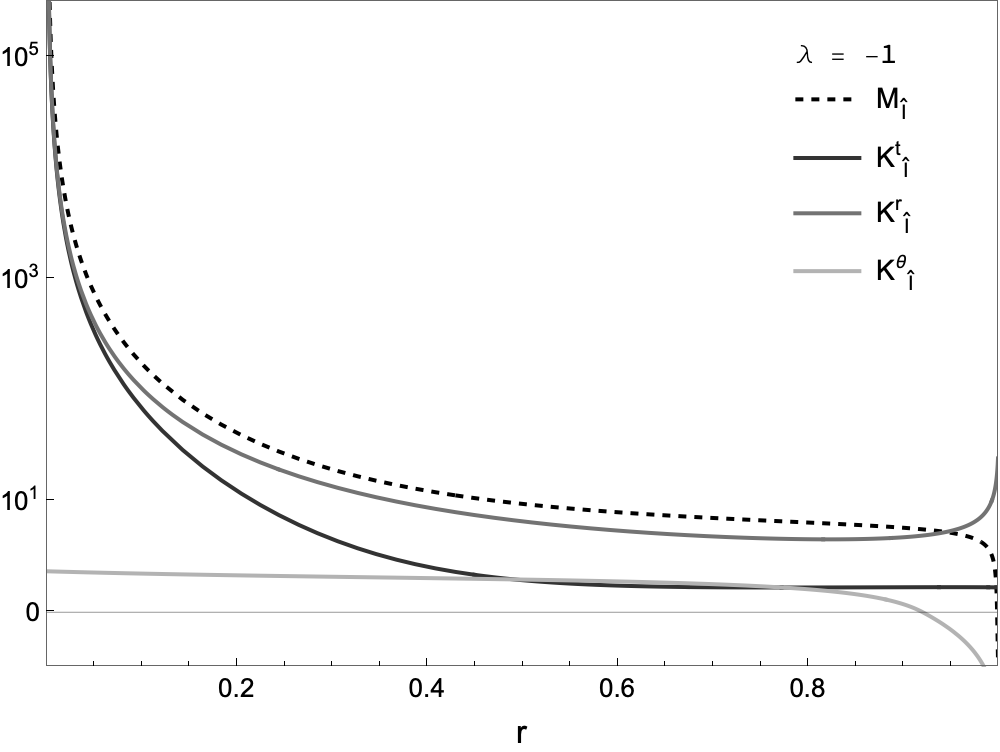}
    
    \includegraphics[width=0.48\linewidth]{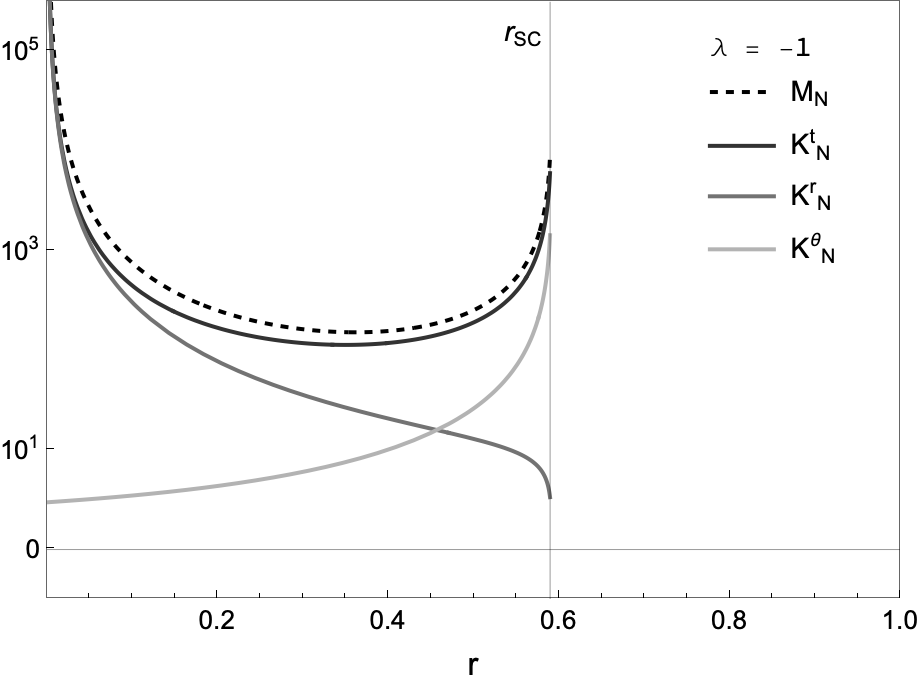}
    \includegraphics[width=0.48\linewidth]{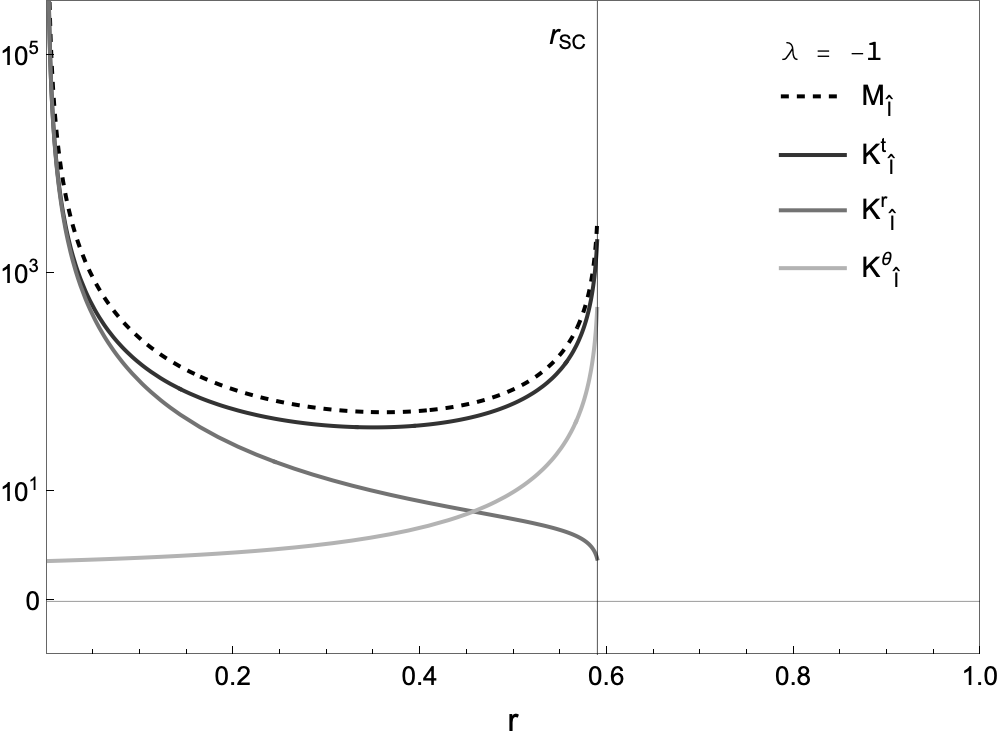}
    \caption{Masses and kinetic coefficients for the $N$ sector (first column) and the $\hat I$ sector (second column) for $\lambda=-1$, evaluated on solutions obtained by integrating from the origin. From top to bottom: $c_4=0.1,\,d_4=-1$; $c_4=1,\,d_4=1$; $c_4=1,\,d_4=-1$. The top and bottom rows correspond to the bottom-left and bottom-right panels of Fig.~\ref{fig:Plot12}.}
    \label{fig:StabilityLambdaMinus1Origin}
\end{figure}

Fig.~\ref{fig:StabilityLambdaMinus1Origin} shows the masses and kinetic coefficients of the $N$ and $\hat I$ perturbations for $\lambda=-1$, evaluated on solutions integrated from the origin. The three rows correspond to $c_4=0.1,\,d_4=-1$ (top), $c_4=1,\,d_4=1$ (middle), and $c_4=1,\,d_4=-1$ (bottom), probing both sides of the condition $c_4^2>1/3$, which controls near-origin stability and the existence of a strong-coupling scale.
For $c_4^2<1/3$ (top row), no strong-coupling radius appears, but ${\cal K}^\theta_N<0$, so the $N$ sector is already unstable near the origin. The $\hat I$ sector remains positive, though both sectors become superluminal at larger radius.
For $c_4^2>1/3$ (middle and bottom rows), the $N$ sector is initially stable. If $d_4>0$ (middle), $r_{\rm SC}<0$ and so the solution spans the entire static patch, but stability is lost at large $r$: ${\cal M}_N$ and ${\cal K}^\theta_N$ turn negative, ${\cal K}^t_N$ approaches zero, ${\cal K}^r_N$ grows, and the radial mode becomes superluminal. If $d_4<0$ (bottom), $r_{\rm SC}\in(0,1)$ and the solution exists only up to this point, but remains well-behaved throughout: all kinetic coefficients are positive and propagation is subluminal. This realizes stable screening in $0<r<r_{\rm SC}$. 
Using Eqs.~\eqref{eq:c2c4d4} and~\eqref{eq:Mpi}, for $\lambda=-1$ we find $d_4 \sim M c_4 (\dd P/\dd \pi^2)_{r=0}$, thus relating $d_4$ to $M$ and $c_4$. In the case of the stable solution just found, we have that $d_4$ and $c_4$ are discordant, thus the coupling with matter must satisfy
\be
(\dd P/\dd \pi^2)_{r=0} < 0,
\ee
so our solution can be accommodated by a coupling of the form $P(\pi^2) = - \pi^2$, or $P(\pi^2) = 1/\pi^2$.

\subsubsection{Stability for $\lambda = -1$, integrating from the horizon}
\begin{figure}[H]
    \centering
    \includegraphics[width=0.48\linewidth]{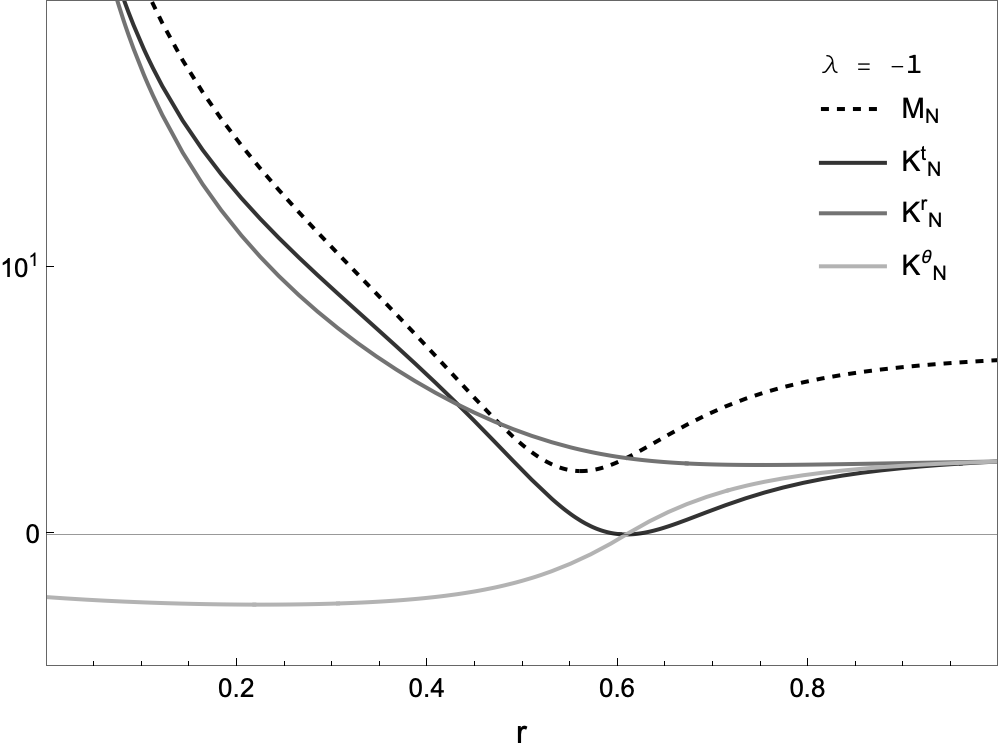}
    \includegraphics[width=0.48\linewidth]{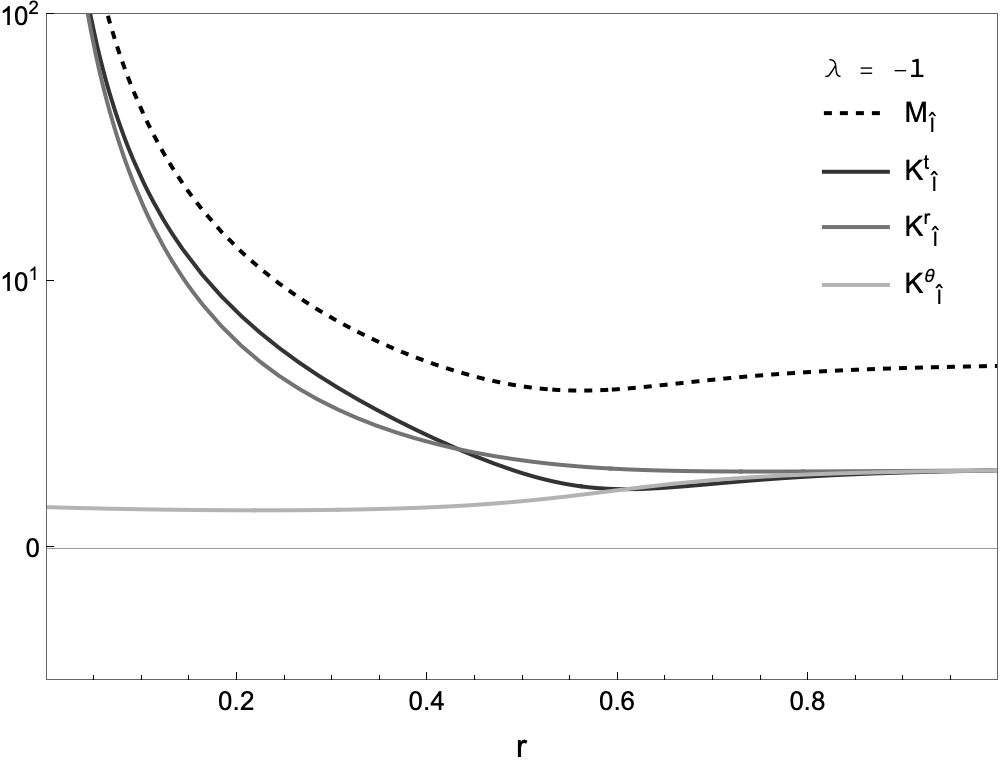}
    \includegraphics[width=0.48\linewidth]{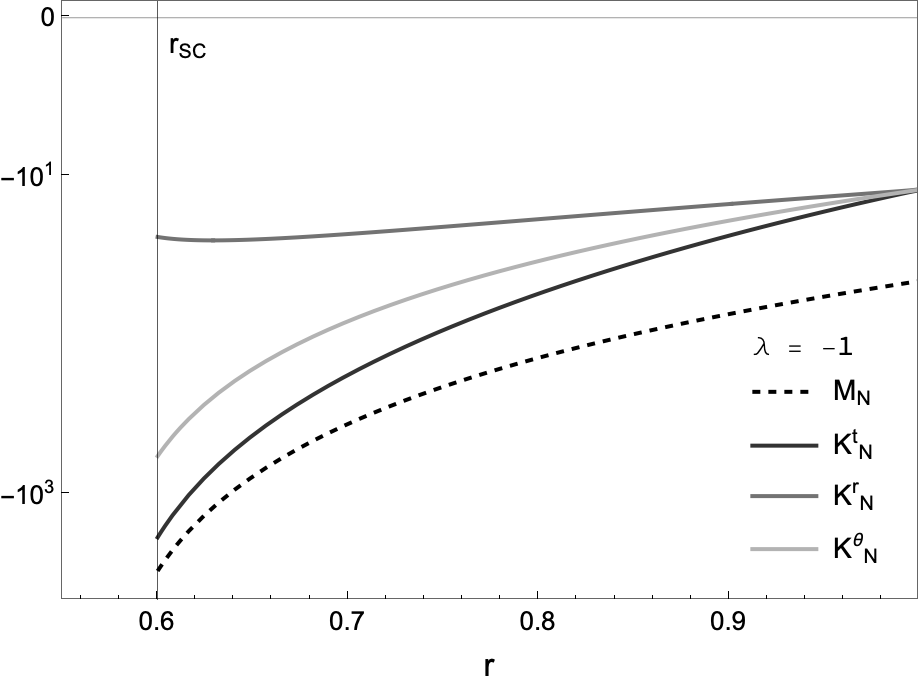}
    \includegraphics[width=0.48\linewidth]{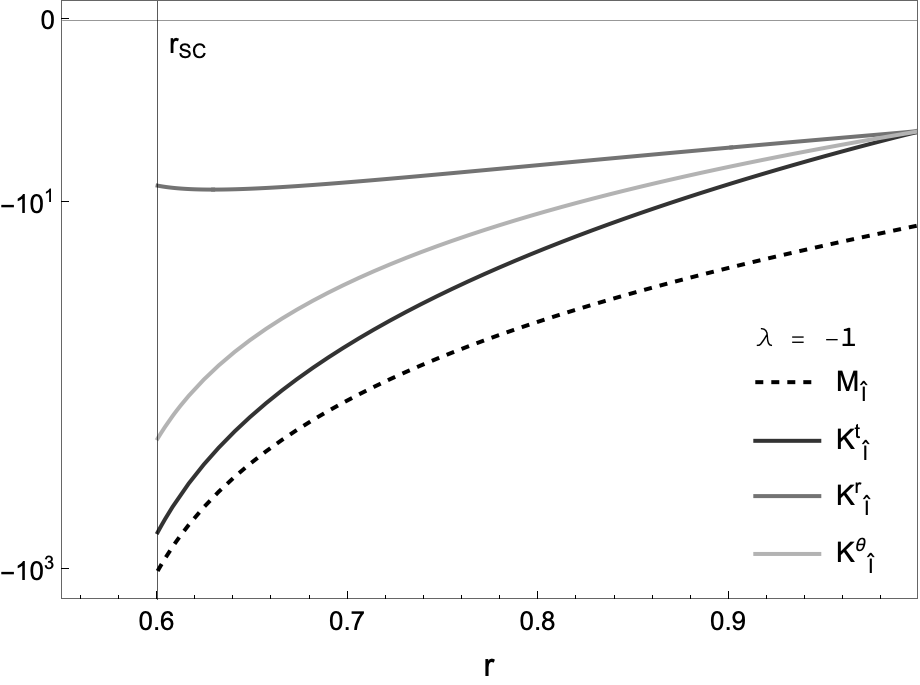}
    \caption{Masses and kinetic coefficients for the $N$ sector (first column) and the $\hat I$ sector (second column) for $\lambda=-1$, evaluated on solutions obtained by integrating from the horizon. The top row corresponds to $c_2=-0.1$ (bottom-left panel of Fig.~\ref{fig:Plot34}), with no strong-coupling radius in the domain. The bottom row corresponds to $c_2=-1$ (bottom-right panel of Fig.~\ref{fig:Plot34}), where a strong-coupling radius is present, and the solution exists only for $r\in(r_{\rm SC},1)$.}
    \label{fig:StabilityLambdaMinus1Horizon}
\end{figure}

Fig.~\ref{fig:StabilityLambdaMinus1Horizon} shows the coefficients of the quadratic action for $\lambda=-1$, evaluated on solutions obtained by integrating from the horizon, using $d_2 = 0$. For $c_2=-0.1$ (top row), $1-12c_2^2>0$, so $B>0$ near the horizon and no strong-coupling point appears. The perturbations are stable there, but this does not extend across the static patch: as the solution is continued inward, ${\cal K}^\theta_N$ becomes negative, and the radial and angular modes are superluminal over an intermediate range.
For $c_2=-1$ (bottom row), $1-12c_2^2<0$, so $B<0$ near the horizon and $B>0$ near the origin, implying a zero in between and hence a strong-coupling radius. The solution is then defined only for $r\in(r_{\rm SC},1)$, where $B<0$, and the perturbations are unstable throughout the domain, consistent with Eqs.~\eqref{eq:KNT}--\eqref{eq:KNTheta}.

\subsection{Goldstone modes and their kinetic terms}
Unlike the flat-space multi-Galileon, where the Galilean symmetry forbids a potential, the dS galileon theory has the potential
\be
    V(\pi)\equiv -\frac{1}{\sqrt{-g}}\,{\cal L}\Big|_{\pi^I=\mathrm{const}}  =- 2H^2\lp \pi^2-\frac{3H^4\lambda}{2}\,\pi^4\rp \,,
\ee
where we have reintroduced units to show explicitly that the potential vanishes for $H=0$.
For $\lambda\le 0$, the potential has a unique maximum at $\pi^2=0$. For $\lambda>0$, it takes the standard symmetry-breaking form: the origin becomes a local maximum, while the minima  lie on the $(N-1)$-sphere defined by
\be
\pi^2=\frac{1}{3H^4\lambda}\,.\label{symbreakse}
\ee
The vacuum manifold therefore breaks $\mathfrak{so}(N)\to\mathfrak{so}(N-1)$, producing $N-1$ Goldstone modes.

One can evaluate the quadratic action of Eq.~\eqref{eq:QuadLagrangianAligned} around the constant symmetry-preserving vev, $\pi^2_0 = 0$, and symmetry breaking vev  $\pi^2_0 = {1\over  3 H^4 \lambda}$. Aligning the background along the last direction in field space, $\pi_0^I=\pi_0\,\delta^I_N$, we find 
\be
    \frac{1}{\sqrt{-g}}\,{\cal L}^{\rm tot}_{\rm quad} =  \llp \lp 1-3H^4\lambda\pi_0^2\rp \delta^{IJ} +6H^4\lambda\,\pi_0^I\pi_0^J \rrp \lp - \frac12 \phi_I^{\ \mu}\phi_{J\mu} + 2 H^2\phi_I\phi_J\rp \,.
\ee
From this expression it is clear that, for the symmetric vacuum, $\pi_0^I=0$, 
\be
    \frac{1}{\sqrt{-g}}\,{\cal L}^{\rm tot}_{\rm quad} =  -\frac12 \phi^{I\mu}\phi_{I\mu} + 2 H^2\phi^I\phi_I\,,
\ee
namely there exist $N$ ghostly scalars with mass $m^2=-4H^2$, while for the  symmetry-breaking vacuum, $\pi_0^I= \delta^I_N / \sqrt{3H^4\lambda}$, the term proportional to $\delta^{IJ}$ vanishes, leaving only 
\be
    \frac{1}{\sqrt{-g}}\,{\cal L}^{\rm tot}_{\rm quad} = - \phi_N^{\ \mu}\phi^{N}_{\mu} + 4H^2\phi^N\phi_N  \,.
\ee
Thus, only the radial mode $\phi^N$ survives in the quadratic action, and the $N-1$ Goldstone modes are absent altogether. The Goldstone excitations have vanishing quadratic kinetic terms about the symmetry-breaking vacuum, and are therefore infinitely strongly coupled. At this point, one cannot reliably identify the perturbations as the correct low energy degrees of freedom of the theory and the effective description ceases to be valid.
An analogous feature was already observed in~\cite{Garoffolo:2025igz} for the full DBI-like theory, where it was attributed to the clash between the dS Galileon shifts~\eqref{eq:GalileonShiftsStatic} and the non-linearly realized  internal symmetry~\eqref{eq:soSymmetry}. 
Indeed, expanding around the symmetry-breaking vev, the broken generators of $\mathfrak{so}(N)$ act as
\bea
    \delta_{\Omega_{\hat{J} N}}\phi^{\hat I}&=&-\frac{1}{\sqrt{3H^4\lambda}}\,\delta^{\hat I}_{\hat J}\,, \nn\\
\delta_{\Omega_{JN}}\phi^N&=&0\,.
\eea
Thus, the Goldstone fields transform by a constant shift, which forbids a mass term.  At the same time, the dS Galileon symmetry fixes the mass of fluctuations around a constant background to $m^2=-4H^2$. These two requirements are incompatible, thus the Goldstone kinetic terms must vanish. By contrast, the radial mode $\phi^N$ does not transform under the broken shift and is therefore allowed to retain both a kinetic term and a mass term.
We note that this phenomenon has been observed in massive gravity as well: linearizing around a specific background a non-linear theory (such as dRGT massive gravity) leads to the vanishing of kinetic terms for scalar and vector perturbations, which are therefore classically strongly coupled~\cite{Boulware:1972yco,Gumrukcuoglu:2011zh,Goon:2014ywa}.

Our analysis of radial solutions on a dS background was initially motivated by the question of whether a nontrivial profile can interpolate between the vacua $\pi^2=0$ and $\pi^2=1/(3H^4\lambda)$ while restoring the Goldstone kinetic terms away from the symmetry-breaking vacuum. 
This would be an instanton-like solution~\cite{Coleman:1977py,Callan:1977pt}, but in the present case we find that no such solution exists. First, the symmetry-breaking vacuum exists only for $\lambda>0$, which is precisely the regime in which the radial equation becomes stiff (Sec.~\ref{sec:stiff}). The solution cannot be continued beyond the strong-coupling radius, so no globally regular configuration  consistent with our ansatz \eqref{piansatze} can connect the two vacua between the origin and the horizon. More generally, even locally such an interpolation is not possible: the Goldstone mass and kinetic terms (Eqs.~\eqref{eq:MassIterm} and~\eqref{eq:KineticIterm}, respectively) depend on $\pi_0$, $\pi_0'$, and $\pi_0''$, and vanish at the symmetry-breaking vacuum only if $\pi_0'=\pi_0''=0$. Thus, it is not sufficient that the solution $\pi(r)$ interpolates between the vacua, it has to do so with a zero derivative. However, for a second-order equation fixing $\pi_0$ and $\pi_0'$ at a point uniquely determines the solution. Thus, imposing vacuum values with vanishing derivative therefore selects the trivial constant branch. We conclude that no nontrivial radial solution in the dS static patch interpolates between the two vacua while restoring the Goldstone kinetic terms away from the symmetry-breaking vacuum.

Nevertheless, we see from the plots in Fig.~\ref{fig:StabilityLambdaMinus1Horizon} that the kinetic terms for the would-be Goldstone modes are non-vanishing around these radial profiles.  This confirms that the vanishing of the kinetic terms in the constant symmetry-breaking solution \eqref{symbreakse} is an artifact of that infinitely strongly coupled solution, and that the kinetic terms, and thus the strong coupling scale, become finite around more realistic non-trivial solutions.

\section{Anti-de Sitter extension}
\label{sec:AdS}

To separate horizon effects from genuine curvature effects, we now consider the same multi-Galileon system on an Anti-de Sitter (AdS) background (generalizing the single field model considered in~\cite{Burrage:2011bt,Goon:2011qf,Goon:2011uw}).
In this case, the spacetime is still curved, so the scalar sector continues to admit a nontrivial potential, but there is no cosmological horizon truncating the radial domain: the static patch metric on AdS is
\be\label{eq:MetricStaticPatchAdS}
    \dd s^2=-\lp 1+ r^2 / L^2 \rp \dd t^2+\frac{\dd r^2}{1+ r^2/ L^2  }+r^2\dd\Omega_2^2\,,
\ee
with $r\in (0,\infty)$ and $L$ the standard AdS radius.
The AdS theory is obtained formally from the dS one by the replacement $H^2 \to - 1 / L^2$, such that the  radial equation of motion has the same structure as Eq.~\eqref{eq:RadialEOM}, but with $\{A, B\} $ replaced with 
\begin{align}
    B^{\rm AdS} &= 2 \lp 1 + \frac{r^{2}}{L^2} \rp \llp  r^{2} - 3\lambda \lp \lp 1+\frac{r^{2}}{L^2} \rp \pi' - \frac{r} {L^2}\,\pi \rp \lp \lp 1+  \frac{3 r^{2}}{L^2} \rp \pi' -\frac{3 r }{L^2}\,\pi \rp \rrp \,,\label{eq:BAdSdef}\\
    A^{\rm AdS} &= 4 \lp \frac{2 r}{L^2} \,\pi+ \lp 1 +  \frac{2 r^{2}}{L^2} \rp \pi' \rp \llp r + \frac{3 \lambda}{L^2} (\pi-r\pi')\lp \lp 1 + \frac{r^{2}}{L^2} \rp \pi' - \frac{r} {L^2} \pi\rp  \rrp \,.
\label{eq:AAdSdef}
\end{align}
The potential becomes
\begin{equation}
    V^{\rm AdS}(\pi) = \frac{4}{L^2}\lp \pi^2-\frac{3\lambda}{2 L^4}\pi^4\rp\,,
\end{equation}
and thus the critical points occur at the same field values as in dS, but their character is reversed: points that were a minimum in dS become a maximum in AdS, and vice versa.

We can now repeat the same steps as in the dS analysis: we determine the asymptotic solutions close to and far from the source and perform a stiffness analysis before numerically solving the equations.
The asymptotic radial solutions are
\begin{align}
    \pi^{\rm AdS}_{(2)}(r) &= \frac{c^{\rm AdS}_2}{r}\lp 1 + \frac{3 r^2}{L^2}\rp  +\frac{d^{\rm AdS}_2}{ r} \llp \frac{3 r}{L}+\lp 1 + \frac{3 r^2}{L^2}\rp  \,\tan^{-1} \lp \frac{r}{L} \rp\rrp \,, \label{eq:pi2soln_AdS}
    \\
    \pi^{\rm AdS}_{(4)}(r) &= \frac{c^{\rm AdS}_4}{L} + \frac{d^{\rm AdS}_4}{L^2}\,r\,, \label{eq:pi4soln_AdS}
\end{align}
where $\{ c^{\rm AdS}_2, d^{\rm AdS}_2, c^{\rm AdS}_4, d^{\rm AdS}_4\}$ are dimensionless integration constants.
The key qualitative difference from dS is that the hyperbolic arctangent is replaced by the ordinary arctangent in the $\pi_{(2)}$ asymptotic solution. The second branch therefore remains finite at all radii, and there is no analog of the horizon regularity condition that fixes one integration constant. In this respect, AdS is closer to flat space than dS, although the large-radius behavior still differs from the Minkowski case: the quadratic solution does not scale as $1/r$. Instead, for $r\gg 1$,
\begin{equation}
    \pi^{\rm AdS}_{(2)}(r\gg 1) \approx \lp \frac{3 r}{L^2}+\frac{1}{ r}\rp \lp c^{\rm AdS}_2+\frac{\pi}{2}d^{\rm AdS}_2\rp -\frac{4\,d^{\rm AdS}_2 L^3}{15 r^4}\,, \label{eq:pi2soln_AdS_bigR}
\end{equation}
which has both a growing and a decaying branch. Only by imposing the relation
\begin{equation}
2c^{\rm AdS}_2=-\pi d^{\rm AdS}_2\,,
\end{equation}
can we remove the growing branch, obtaining an asymptotic decay $\sim 1/r^4$. 

The stiffness analysis in AdS proceeds in the same way as the one in dS. Evaluating $B^{\rm AdS} $ on the two asymptotic branches gives
\begin{align}
    B^{\rm AdS}_{r\sim L} &= 4L^2 \llp 1 + 12\frac{\lambda}{L^6}c^{\rm AdS}_2d^{\rm AdS}_2 + 3\frac{\lambda}{L^6}(d^{\rm AdS}_2)^2 + 3\pi\frac{\lambda}{L^6}(d^{\rm AdS}_2)^2 \rrp\,, \\
    B^{\rm AdS}_{r\sim 0} &= -6 \frac{\lambda }{L^6} \lp c^{\rm AdS}_4\rp^2\,.
\end{align}
In contrast to dS, where for $\lambda>0$ the sign change of $B$ is unavoidable, in AdS the sign of $B^{\rm AdS}$ depends sensitively on the boundary data. For either sign of $\lambda$, some choices keep $B^{\rm AdS}$ definite in sign, while others make it cross zero. In the latter case, the local analysis shows that the obstruction cannot be removed by requiring $A^{\rm AdS}=B^{\rm AdS}=0$, so the equation becomes stiff and a strong-coupling scale emerges, as in dS. 
Introducing
\be\label{eq:AdSuv}
    u(r)\equiv \frac{1}{L^2} \lp \pi-r\pi' \rp\,,\qquad v(r)\equiv - \frac{r}{L^2}\pi+\lp 1 + \frac{r^2}{L^2}\rp\pi'\,,
\ee
we can write $B^{\rm AdS}$ and $A^{\rm AdS}$ of Eqs.~\eqref{eq:BAdSdef} and~\eqref{eq:Adef}, as
\begin{align}
    B^{\rm AdS} &= 2\lp 1 + \frac{r^2}{L^2}\rp \Big[r^2-3\lambda\,v(v - 2ur)\Big]\,,\\
    A^{\rm AdS} &= 4(v- ur)\lp r+3\lambda uv\rp \,.
\end{align}
These expressions allow us to study whether $A^{\rm AdS}, B^{\rm AdS} $ may vanish at the same radial point, so to possibly avoid a strong coupling scale.
$A^{\rm AdS}=0$ implies either $v - ur =0$ or $r+ 3\lambda uv=0$.
Substituting these conditions back into $B$ yields
\begin{align}
    B^{\rm AdS}\Big|_{ur-v=0} &= 2\lp 1 + \frac{r^2}{L^2}\rp\lp r^2+3\lambda {v}^2\rp \,,\\
    B^{\rm AdS}\Big|_{r+3\lambda uv=0} &= -2\lp 1 + \frac{r^2}{L^2}\rp\lp r^2+3\lambda {v}^2\rp \,.
\end{align}
Analogously to the dS case, for $\lambda > 0$ the quantity in parentheses remains strictly positive for all real values of $v$. In this case, however, there is no corresponding restriction on the domain of the radial coordinate $r$. Moreover, for $\lambda > 0$ it follows that the conditions $A^{\rm AdS} = B^{\rm AdS} = 0$ cannot be satisfied simultaneously. Consequently, at points where $B^{\rm AdS} = 0$, the differential equation becomes stiff, and the radial solution cannot be extended smoothly across such points.
In contrast, for $\lambda < 0$, it becomes possible to satisfy $A^{\rm AdS} = B^{\rm AdS} = 0$, in direct analogy with the de Sitter case.

The strong coupling radius develops at 
\be
    \frac{r^{\rm AdS}_{\rm SC}}{L} = -\frac{6 (\lambda/ L^6)  c^{\rm AdS}_4 d^{\rm AdS}_4 \pm \sqrt{3 (\lambda/ L^6) (d^{\rm AdS}_4)^2\lp 1+3 (\lambda/ L^6) (c^{\rm AdS}_4)^2  \rp }}{1-9 (\lambda/ L^6)  (c^{\rm AdS}_4)^2}\,,
\ee
which is the opposite sign compared to that in the dS case.

The AdS quadratic action can be found by also substituting $H^2 \to - 1 / L^2$ in Eq.~\eqref{eq:QuadLagrangianAligned}.  The kinetic coefficients of the $N$ modes take the simple form
\begin{align}
    {\cal K}^{t, {\rm AdS}}_{N} &= \frac{2(1+r^2/L^2)}{r^2}\,\frac{\bigl(r^2+3\lambda v^2\bigr)^2}{B^{\rm AdS}}\,, \label{eq:KNTAdS}\\
    {{\cal K}^{r, {\rm AdS}}_{N}} &= \frac{B^{\rm AdS}}{2r^2(1+r^2/L^2)}\,, \label{eq:KNRAdS}\\
    {\cal K}^{\theta, {\rm AdS}}_{N} &= 2(1+r^2/L^2)\,\frac{(1+3\lambda u^2)(r^2+3\lambda v^2)}{B^{\rm AdS}} \label{eq:KNThetaAdS}\,,
\end{align}
where now $u,v$ are the AdS variables in Eq.~\eqref{eq:AdSuv}, and $B^{\rm AdS}$ is defined in Eq.~\eqref{eq:BAdSdef}.
The mass coefficient is given by 
    \begin{align}
        {\cal M}^{\rm AdS}_N &=  - \frac{4 (1 + r^2/L^2 )}{B^{AdS} L^2 r^2 } \lp 2 r^4 + 3 r^2 (v + u r)^2 \lambda + 9\lambda^2  v^2 (v^2 - 2 u v r + 3 u^2 r^2) \rp\,,\label{eq:MassNtermnoH}
\end{align}
These expressions show, once again, that for $\lambda >0$, the signs of the kinetic coefficients are entirely dependent on the sign of $B^{\rm AdS}$, and perturbations are stable only if $B^{\rm AdS} >0$. 
For $\lambda <0$, in addition to the positivity of $B^{\rm AdS} >0$, one must also require $(1+3\lambda u^2)(r^2+3\lambda v^2) > 0$, otherwise the angular mode becomes unstable. 
We can evaluate these coefficients in two different regimes, using the asymptotic branches identified in the AdS case: Eq.~\ref{eq:pi4soln_AdS} near the origin and Eq.~\ref{eq:pi2soln_AdS} as $r \gg 1$. On the near-source branch, the leading asymptotics are the same as those of dS (see Eq.~\eqref{eq:KineticCloseOrigin}), and in the large r limit where $r \gg 1$ we find 
\be
    \lim_{r \gg L} \lp {\cal K}^{t, {\rm AdS}}_N\,, {\cal K}^{r, {\rm AdS}}_N \,,{\cal K}^{\theta, {\rm AdS}}_N \rp = 1 \,,
\ee
independently of the boundary condition choices of $c_2$ and $d_2$. The corresponding sound speeds are 
\be 
    (c_{N}^{r, {\rm AdS}})^2 = \frac{\lp r^2 - 3 \lambda v (v - 2 r u)\rp^2 }{ \lp r^2+3\lambda v^2 \rp^2}\,,  \qquad  (c_{N}^{\theta, {\rm AdS}})^2 = \frac{(1+3\lambda u^2)r^2}{r^2+3\lambda v^2} \,,
\ee 
which are required to lie in $\in (0,1]$.
The perturbations of the $\hat{I}$ modes are once again trivially related to the $N$ modes via
\be 
{\cal K}^{t, {\rm AdS}}_{\hat I} = \frac{2+{\cal K}^{t, {\rm AdS}}_N}{3} \,, \quad {\cal K}^{r, {\rm AdS}}_{\hat I} = \frac{2+{\cal K}^{r, {\rm AdS}}_N}{3} \,, \quad {\cal K}^{\theta, {\rm AdS}}_{\hat I} = \frac{2+{\cal K}^{\theta, {\rm AdS}}_N}{3} \,, \label{eq:ImodesasNmodesAdS}
\ee 
for the kinetic terms, and 
\be 
{\cal M}^{\rm AdS}_{\hat I} = \frac{-8/L^2+{\cal M}^{\rm AdS}_N}{3} \,, \quad 
\ee 
for the mass.  Thus the stability analysis of the $N$ modes translates to the ${\hat I}$ modes straightforwardly.

\subsection{Numerical analysis}
As for dS, we perform the numerical analysis in dimensionless variables, rescaling the quantities under consideration as
\be \label{eq:rescalingAdS}
    \pi \to \pi/L \,, \qquad r \to r L \,, \qquad \lambda \to  L^6 \lambda\,, \qquad M \to M / L\,.
\ee 

\begin{figure}[H]
    \centering
    \includegraphics[width=0.5\linewidth]{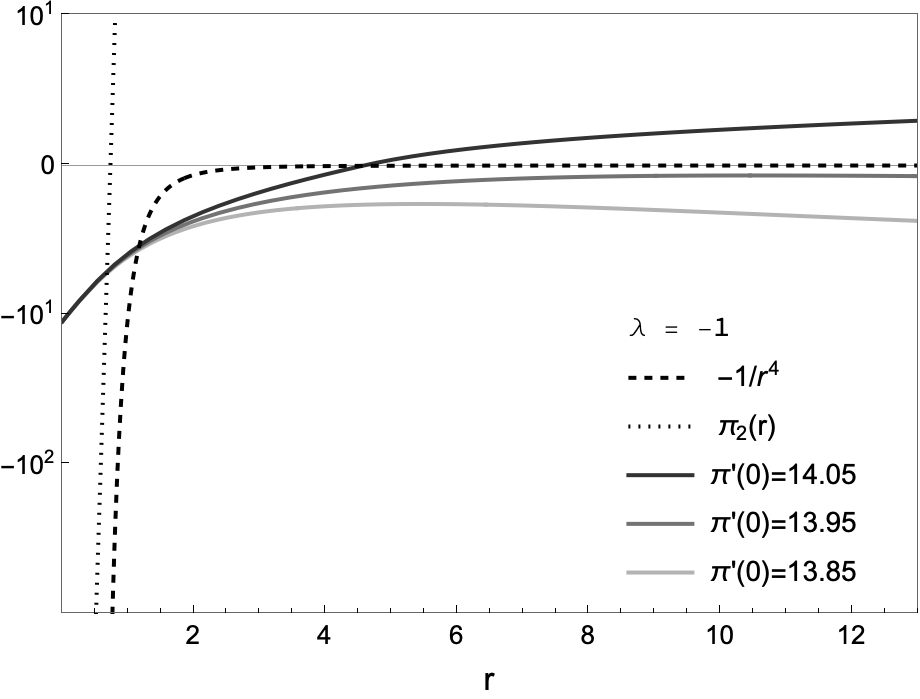}
    \caption{Numerical solutions to the radial equation in AdS, for boundary conditions $\pi(0)=-10$ and $\pi'(0)=\{13.85,\,13.95,\,14.05\}$. The dotted black line shows the near-origin analytic solution $\pi^{\rm AdS}_{(4)}$, while the solid black curve indicates the asymptotic $-1/r^4$ behavior.}
    \label{fig:piAds}
\end{figure}

An example of the numerical behavior is shown in Fig.~\ref{fig:piAds}, where we solve the AdS equation with boundary condition $\pi(0)=-10$ and  $\pi'(0)=\{13.85,\,13.95,\,14.05\}$. The dotted curve is the near-origin analytic branch $\pi^{\rm AdS}_{(4)}$, while the black curve indicates the asymptotic $-1/r^4$ behavior. Small changes in the initial derivative lead to qualitatively different large-radius profiles, showing that the asymptotic branch is highly sensitive to the initial data. This reflects the fact that, in AdS, one can tune solutions to approach a vacuum only asymptotically, rather than at finite radius as in dS.

\section{Conclusions}

In this paper we have studied static, spherically symmetric solutions of the multi-field Galileon theory on de Sitter space with \(\mathfrak{so}(N)\)-invariant matter couplings. After aligning the background along one field-space direction, the system reduces to a single nonlinear radial equation, which we analyzed both analytically and numerically.

Near the source, the quartic interaction generates a nonlinear screened branch, while at larger radius the quadratic term dominates, defining a Vainshtein scale \(r_{\rm V}\). In de Sitter space, however, this is not sufficient to characterize viable solutions. The radial equation can become singular at a finite radius \(r_{\rm SC}\), which sets the boundary of the domain where the effective description is valid. Screening must therefore be understood together with the requirement of regularity.
Our analysis shows that stable, subluminal screened solutions do exist, both for profiles integrated from the origin and from the horizon, but only within a restricted interval set by \(r_{\rm SC}\). The presence of this strong-coupling scale is essential: it determines the domain over which the solution remains well defined and allows a consistent screened regime to emerge. In this sense, screening in de Sitter is intrinsically tied to the finite region bounded by \(r_{\rm SC}\), rather than to specific asymptotic conditions, and its existence is controlled by the interplay between the Vainshtein and strong-coupling scales.
We also briefly considered Anti-de Sitter space, where the absence of a cosmological horizon removes the de Sitter obstruction. In that case, the asymptotic behavior is not fixed at finite radius and depends sensitively on the boundary conditions, allowing solutions that approach a vacuum only asymptotically.

Finally, we examined whether a radial profile can interpolate between the symmetric and symmetry-breaking vacua while restoring the Goldstone kinetic terms. In de Sitter this is not possible. The symmetry-breaking vacuum lies in the branch where singular behavior arises, and imposing the vacuum value with vanishing derivatives fixes the trivial constant solution. There is therefore no nontrivial profile that connects the two vacua and resurrects the Goldstone dynamics.

Overall, curvature qualitatively modifies the flat-space picture: screening can occur in a stable and controlled way, but only within a finite region bounded by the strong-coupling scale, which plays a central role in both the background dynamics and the behavior of perturbations.

\section*{Acknowledgments}
We would like to thank Hayden Lee for useful discussions during the preparation of this manuscript.
A.G. is supported by funds provided by the Center for Particle Cosmology at the University of Pennsylvania. M.G.~is supported by the U.S. Department of Energy (DOE), Office of Science, Office of Advanced Scientific Computing Research, Department of Energy Computational Science Graduate Fellowship under Award Number DE-SC0023112. 
The work of M.T. is supported in part by DOE (HEP) Award DE-SC0013528. 

\appendix

\section{Quadratic action}
\label{app:QuadraticAction}
Expanding the action~\eqref{eq:Action} to second order in the field perturbation, $\pi^I = \pi^I_0 + \phi^I$, we obtain 
\begin{align}
    \frac{1}{\sqrt{-g}}\,{\cal L}^{\rm tot}_{\rm quad} =& +\phi^I\phi_I \llp  4H^2-12H^6\lambda \pi_0^J \pi_{0J} +3H^4\lambda \pi_0^{J\mu}\pi_{0 J\mu} \rrp + \nonumber\\
    &+\phi_I\phi_J \llp  -24H^6\lambda \pi^I_0 \pi^J_0  +6H^4\lambda \pi_0^{J \mu}\pi^I_{0\mu} \rrp + \nonumber\\
    &+\phi^I\phi_{I \mu} \llp  12H^4\lambda \pi_0^J \pi_{0 J}^\mu \rrp + \nonumber\\
    &+\phi_I\phi_{J\mu} \llp  12 H^4\lambda \lp \pi^I_0 \pi^{J\mu}_0 + \pi^J_0 \pi^{I\mu}_0 \rp  +4H^2\lambda \pi_0^{J\mu} \Box \pi^I_0 - 4H^2\lambda \pi_0^{J\nu} \pi^{I\mu}_{0 \ \nu} \rrp + \nonumber\\
    &+\phi^{I\mu}\phi_{I\mu}\llp -1+3H^4\lambda \pi^J_0 \pi_{0J} +2H^2\lambda \pi^J_0 \Box \pi_{0J} +\frac52 H^2\lambda \pi^{J \nu}_0 \pi_{0J \nu}\rrp + \nonumber\\
    &+\phi_{I}^\mu\phi_{J \mu} \llp  6H^4\lambda \pi^I_0 \pi^J_0 -H^2\lambda \pi_0^{I \nu} \pi^J_{0 \nu} \rrp \nonumber\\
    &+\phi^{I}_{\ \nu} \phi_{I \mu}\llp -2H^2\lambda \pi_{0 J} \pi^{J\mu\nu}_{0} - H^2\lambda \pi_{0 J}^\mu \pi^{J\nu}_{0}  \rrp + \nonumber\\
    &+\phi_{I\mu}\phi_{J\nu}\llp 5H^2\lambda \pi_0^{I\mu}\pi_0^{J\nu} -H^2\lambda \pi_0^{J\mu}\pi_0^{I\nu} +\lambda \pi^{I\nu}_{0 \rho} \pi_0^{J\mu\rho}  -\lambda \pi_0^{I\mu\nu} \Box \pi^J_0\rrp + \nonumber\\
    &+\phi^I\phi_{I\mu\nu} \llp  -2H^2\lambda \pi_{0 J}^\mu \pi^{J\nu}_{0}  \rrp + \phi^I\Box\phi_I \llp  2H^2\lambda \pi^{J\mu}_{0}  \pi_{0 J \mu}  \rrp + \nonumber\\
    &+\phi_{I\mu}\Box\phi^I \llp  -\lambda  \pi_0^{J\mu\nu} \pi_{0 J \nu} \rrp +\phi_{I \mu} \Box \phi_J \llp 4H^2\lambda \pi^J_0 \pi^{I\mu}_0 - \lambda \pi_{0\nu}^{J} \pi_0^{J\mu\nu} \rrp + \nonumber\\
    &+\phi^{I\mu}\phi_{I\mu\nu}\llp -\lambda \pi_0^{J\nu} \Box \pi_{0J} \rrp + \nonumber\\
    &+\phi_I^{\ \mu}\phi_{J\mu\nu} \llp -4H^2\lambda \pi_0^J \pi_0^{I\nu} -\lambda \pi^{J\nu}_0 \Box \pi^I_0 +2\lambda \pi_{0\rho}^{J} \pi_0^{I \rho \nu}  \rrp + \nonumber\\
    &+\phi^{I}_\mu \phi_{I\rho\nu} \llp  2\lambda \pi_0^{J\nu} \pi_{0J}^{\rho \mu} \rrp  +\phi_{I\mu\nu}\Box\phi_J \llp  -\lambda \pi_0^{I\mu} \pi_0^{J\nu} \rrp  +\phi_{I\rho\mu}\phi_{J\nu}^{\ \ \rho} \llp  \lambda \pi_0^{J\mu}\pi_0^{I\nu} \rrp \,.
\label{eq:LagrangianQuad}
\end{align}
We can further simplify this expression by performing integrations by parts (here we assume that $\pi^I_0 = \pi^I_0 (r)$). We obtain

\begin{align}
    &\frac{1}{\sqrt{-g}}\,{\cal L}^{\rm tot}_{\rm quad} =  - \phi^{I\mu} \phi_{I\mu} + 4 H^2 \phi^I \phi_I + \nonumber  \\
    &~+ \lambda \phi^I \phi_I    \llp - 12 H^6 \pi^2_0  + 3   H^4  \lp \pi_0^{J\mu} \pi_{0 J \mu}  -    \Box \pi^2_0 \rp + H^2 \Box \lp \pi_0^{J\mu} \pi_{0 J \mu} \rp - H^2 \nabla_\mu \nabla_\nu \lp \pi_0^{J \mu} \pi_{0J}^{\nu} \rp \rrp  + \nonumber\\
    &~+ \lambda\phi_I \phi_J   \llp  - 24 H^6 \pi^I_0 \pi^J_0  + 12 H^4 \pi_0^{I\mu} \pi_{0\mu}^J - 6H^4 \Box \lp \pi^I_0 \pi^J_0  \rp + 2 H^2 \lp \pi_{0\mu\nu}^J \pi_0^{I\mu\nu}   -   \Box \pi_0^I \Box \pi_0^J \rp \rrp + \nonumber\\
    &~+ \lambda \phi^{I\mu} \phi_{I\mu} \llp  3  H^4   \pi^2_0 + 2  H^2   \pi_0^J \Box \pi_{0J} + \frac32 H^2   \pi_0^{J\nu} \pi_{0 J \nu}  +\frac12  \nabla_\nu \lp \pi_0^{J\nu }  \Box \pi_{0J} \rp -\frac14  \Box \lp \pi_0^{J\nu} \pi_{0 J \nu} \rp  \rrp + \nonumber\\
    &~+\lambda\phi_{I\mu} \phi^{I}_\nu  \llp - 2 H^2  \pi_{0J}\pi_0^{J\mu\nu}    - \pi_0^{J\mu\nu}  \Box \pi_{0J} + \pi_0^{J\mu\rho} \pi_{0 J \rho}^{\ \ \ \nu}  \rrp + \nonumber\\
    &~+ \lambda \phi_I^{\mu} \phi_{J \mu}  \llp 6 H^4 \pi^I_0 \pi^J_0   + 2 H^2 \lp \pi^I_0 \Box \pi^J_0  + \pi^J_0  \Box \pi^I_0   \rp + \Box \pi^I_0 \Box \pi^J_0  - \pi_0^{I\nu\rho} \pi_{0\nu\rho}^{J} \rrp + \nonumber\\
    &~+ \lambda  \phi_{I\mu} \phi_{J\nu} \llp 4 H^2  \pi_0^{I\mu} \pi_0^{J\nu} - 6 H^2  \pi_0^{J\mu} \pi_0^{ I \nu}  -  4 H^2   \pi_0^I \pi_0^{J\mu\nu}  + \pi_0^{J\nu\rho} \pi_{0\rho}^{I\mu} +  \pi_0^{I\nu\rho} \pi_{0\rho}^{J\mu}  \rrp + \nonumber\\
    &~+ \lambda  \phi_{I\mu} \phi_{J\nu} \llp  - \pi_0^{J \mu\nu} \Box \pi_0^I  - \pi_0^{I \mu\nu} \Box \pi_0^J \rrp
 \label{eq:LagrangianQuadparts4}
\end{align}
with $\pi^2_0 \equiv \pi_{0}^J \pi_{0 J} $.

\subsection{Coefficients of the quadratic action}
\label{app:CoefficientQuadraticAction}
Starting from Eq.~\eqref{eq:LagrangianQuadparts4}, one can evaluate the parentheses by setting $\pi^I_0 = \delta^I_N \pi_0(r)$ and derive Eq.~\eqref{eq:QuadLagrangianAligned}. Here, we have reintroduced units to see explicitly how the dS case differs from the flat space case.  
The coefficients $\{{\cal M}_N, {\cal M}_{\hat I} \}$ of the zero derivative mass-like terms proportional to $\phi^I \phi_I$ or $\phi_I \phi_J$ in Eq.~\eqref{eq:QuadLagrangianAligned} are 
\begin{align}
    {\cal M}_N&=4H^2- 36H^6\lambda \pi_0^2 -\frac{36H^4\lambda}{r}\lp 1-2H^2r^2\rp \pi_0\pi_0' -18H^4\lambda(1-H^2r^2)\pi_0\pi_0''\nonumber\\
    &\quad-\frac{6H^2\lambda}{r^2}\lp 1-6H^2r^2+6H^4r^4\rp \pi_0'^2 -\frac{12H^2\lambda}{r}\lp 1-\frac52H^2r^2+\frac32H^4r^4\rp  \pi_0'\pi_0''\,,\label{eq:MassNterm}\\
    {\cal M}_{\hat I}&=4H^2-12H^6\lambda\pi_0^2 -\frac{12H^4\lambda}{r}\lp 1-2H^2r^2\rp \pi_0\pi_0' -6H^4\lambda(1-H^2r^2)\pi_0\pi_0''\nonumber\\
    &\quad-\frac{2H^2\lambda}{r^2}\lp 1-6H^2r^2+6H^4r^4\rp \pi_0'^2 -\frac{4H^2\lambda}{r}\lp 1-\frac52H^2r^2+\frac32H^4r^4\rp  \pi_0'\pi_0''\, .\label{eq:MassIterm}
\end{align}
The kinetic terms are derived from  all the terms with two derivatives in Eq.~\eqref{eq:LagrangianQuadparts4}. Compatible with the symmetries, they are diagonal
as expressed in Eq.~\eqref{eq:KineticNterm} and~\eqref{eq:KineticIterm}, with components
\begin{align}
    {\cal K}^t_N&=1-9H^4\lambda\pi_0^2-\frac{12H^2\lambda}{r}\lp 1-\frac32H^2r^2\rp \pi_0\pi_0'-6H^2\lambda(1-H^2r^2)\pi_0\pi_0''\nonumber\\
    &\quad-\frac{3\lambda}{r^2}\lp 1-4H^2r^2+3H^4r^4\rp \pi_0'^2-\frac{6\lambda}{r}\lp 1-2H^2r^2+H^4r^4\rp \pi_0'\pi_0''\,,\label{eq:KineticNtterm}\\ \nonumber \\
    {\cal K}^r_N&=1-9H^4\lambda\pi_0^2-\frac{12H^2\lambda}{r}\lp 1-\frac32H^2r^2\rp \pi_0\pi_0'-\frac{3\lambda}{r^2}\lp 1-4H^2r^2+3H^4r^4\rp \pi_0'^2\label{eq:KineticNrterm}\\\nonumber \\
    {\cal K}^\theta_N&=1-9H^4\lambda\pi_0^2-\frac{6H^2\lambda}{r}\lp 1-3H^2r^2\rp \pi_0\pi_0'-6H^2\lambda(1-H^2r^2)\pi_0\pi_0''\nonumber\\
    &\quad+6H^2\lambda\lp 1-\frac32H^2r^2\rp \pi_0'^2-\frac{3\lambda}{r}\lp 1-3H^2r^2+2H^4r^4\rp \pi_0'\pi_0''\,,\label{eq:KineticNthetaterm}
\end{align}
for the modes aligned with the background, and 
\begin{align}
    {\cal K}^t_{\hat I}&=1-3H^4\lambda\pi_0^2-\frac{4H^2\lambda}{r}\lp 1-\frac32H^2r^2\rp \pi_0\pi_0'-2H^2\lambda(1-H^2r^2)\pi_0\pi_0''\nonumber\\
    &\quad-\frac{\lambda}{r^2}\lp 1-4H^2r^2+3H^4r^4\rp \pi_0'^2-\frac{2\lambda}{r}\lp 1-2H^2r^2+H^4r^4\rp \pi_0'\pi_0''\,,\\\nonumber \\
    {\cal K}^r_{\hat I}&=1-3H^4\lambda\pi_0^2-\frac{4H^2\lambda}{r}\lp 1-\frac32H^2r^2\rp \pi_0\pi_0'-\frac{\lambda}{r^2}\lp 1-4H^2r^2+3H^4r^4\rp \pi_0'^2\,,\\\nonumber \\
    {\cal K}^\theta_{\hat I}&=1-3H^4\lambda\pi_0^2-\frac{2H^2\lambda}{r}\lp 1-3H^2r^2\rp \pi_0\pi_0'-2H^2\lambda(1-H^2r^2)\pi_0\pi_0''\nonumber\\
    &\quad+2H^2\lambda\lp 1-\frac32H^2r^2\rp \pi_0'^2-\frac{\lambda}{r}\lp 1-3H^2r^2+2H^4r^4\rp \pi_0'\pi_0''\,,
\end{align}
for the remaining ones, where a prime represents a partial derivative with respect to r.
These relations also show that 
\be
{\cal K}^t_{\hat I} =  \frac{2+{\cal K}^t_N}{3} \,, \qquad {\cal K}^r_{\hat I} =  \frac{2+{\cal K}^r_N}{3} \,, \qquad {\cal K}^r_{\hat I} =  \frac{2+{\cal K}^r_N}{3}\,, 
\ee 
while, focusing on the masses in Eqs.~\eqref{eq:MassNterm} and~\eqref{eq:MassIterm}, one sees the relation
\be 
{\cal M}_{\hat I} =  \frac{8 H^2 + {\cal M}_N}{3}\,.
\ee

\bibliographystyle{utphys}
\bibliography{references.bib}

\end{document}